\documentclass[12pt,preprint]{aastex}
\newcommand{\ea}{et al.}

\def\ijhk{$I-$, $J-$, $H-$ and $K-$band }
\def\izjhk{$I-$, $z^{\prime}-$, $J-$, $H-$ and $K-$band }
\def\iz{$I-$ and $z^{\prime}-$band }
\def\izcap{$I-$ and $z^{\prime}-$Band }
\def\jhk{$J-$, $H-$ and $K-$band }
\def\mjup{$M_{Jup}$}
\def\hk{$H-K$}
\def\jh{$J-H$}
\def\msun{$M_{\odot}$}
\newcommand{\gae}{\lower 2pt \hbox{$\, \buildrel {\scriptstyle >}\over {\scriptstyle
\sim}\,$}}
\newcommand{\lae}{\lower 2pt \hbox{$\, \buildrel {\scriptstyle <}\over {\scriptstyle
\sim}\,$}}

\slugcomment{Accepted for publication in ApJ}

\shorttitle{New Young Brown Dwarfs in OMC 2/3}
\shortauthors{Peterson et al.}

\begin{document}

\title{New Young Brown Dwarfs in the Orion Molecular Cloud 2/3 Region\altaffilmark{1}}

\author{Dawn E. Peterson\altaffilmark{2,3,4}}
\affil{Department of Astronomy, University of Virginia,
    Charlottesville, VA 22904}
\email{dpeterson@cfa.harvard.edu}

\author{S. T. Megeath\altaffilmark{2,3,5}}
\affil{Harvard-Smithsonian Center for Astrophysics, 60 Garden Street, 
    Cambridge, MA 02138}
\email{megeath@physics.utoledo.edu}

\author{K. L. Luhman}
\affil{Department of Physics and Astronomy, Pennsylvania State University, 
    University Park, PA 16802}
\email{kluhman@astro.psu.edu}

\author{J. L. Pipher}
\affil{Department of Physics and Astronomy, University of Rochester, 
    Rochester, NY 14627}
\email{jlpipher@astro.pas.rochester.edu}

\author{J. R. Stauffer}
\affil{Spitzer Science Center, California Institute of Technology, MS 314-6, Pasadena, CA 91125}
\email{stauffer@ipac.caltech.edu}

\author{D. Barrado y Navascu\'{e}s}
\affil{LAEFF-INTA, Apartado 50.727, E-28080 Madrid, Spain}
\email{David.Barrado@laeff.esa.es}

\author{J. C. Wilson}
\affil{Department of Astronomy, University of Virginia,
    Charlottesville, VA 22904}
\email{jcw6z@virginia.edu}

\author{M. F. Skrutskie}
\affil{Department of Astronomy, University of Virginia,
    Charlottesville, VA 22904}
\email{mfs4n@virginia.edu}

\author{M. J. Nelson}
\affil{Department of Astronomy, University of Virginia,
    Charlottesville, VA 22904}
\email{mjn4n@virginia.edu}

\and

\author{J. D. Smith}
\affil{Steward Observatory, 933 N. Cherry Ave.,
    Tucson, AZ 85721}
\email{jdsmith@as.arizona.edu}

\altaffiltext{1}{Some of the observations presented here were obtained at the MMT Observatory, Whipple Observatory, W.M. Keck Observatory, Las Campanas Observatory, Mount Graham International Observatory, and Apache Point Observatory.  The MMT Observatory is a joint facility of the Smithsonian Institution and the University of Arizona.  The W.M. Keck Observatory is operated as a scientific partnership among the California Institute of Technology, the University of California and the National Aeronautics and Space Administration. The W.M. Keck Observatory was made possible by the generous financial support of the W.M. Keck Foundation.  The Apache Point Observatory 3.5-meter telescope is owned and operated by the Astrophysical Research Consortium.  In addition, this publication makes use of data products from the Two Micron All Sky Survey, which is a joint project of the University of Massachusetts and the Infrared Processing and Analysis Center/California Institute of Technology, funded by the National Aeronautics and Space Administration and the National Science Foundation.}
\altaffiltext{2}{Visiting Astronomer, Kitt Peak National Observatory, National Optical Astronomy Observatory, which is operated by the Association of Universities for Research in Astronomy, Inc. (AURA) under cooperative agreement with the National Science Foundation.}
\altaffiltext{3}{Visiting Astronomer at the Infrared Telescope Facility, which is operated by the University of Hawaii under Cooperative Agreement no. NCC 5-538 with the National Aeronautics and Space Administration, Science Mission Directorate, Planetary Astronomy Program.}
\altaffiltext{4}{Present address: Harvard-Smithsonian Center for Astrophysics, 60 Garden Street, Cambridge, MA 02138.}
\altaffiltext{5}{Present address: Department of Physics and Astronomy, University of Toledo, Toledo, OH 43606.}

\begin{abstract}

Forty new low mass members with spectral types ranging from M4$-$M9 have been 
confirmed in the Orion Molecular Cloud 2/3 region.  Through deep, \izjhk 
photometry of a 20$^{\prime}$ $\times$ 20$^{\prime}$ field in OMC 2/3, we 
selected brown dwarf candidates for follow-up spectroscopy.  Low resolution 
far-red and near-infrared spectra were obtained for the candidates, 
and 19 young brown dwarfs in the OMC 2/3 region are confirmed.  They 
exhibit spectral types of M6.5$-$M9, corresponding to approximate masses of 
0.075$-$0.015 $M_{\odot}$ using the evolutionary models of \citet{bcah98}.  At 
least one of these {\it bona fide} young brown dwarfs has strong H$\alpha$ 
emission, indicating that it is actively accreting.  In addition, we confirm 
21 new low mass members with spectral types of M4$-$M6, corresponding to 
approximate masses of 0.35$-$0.10 $M_{\odot}$ in OMC 2/3.  By comparing 
pre-main sequence tracks to the positions of the members in the H-R diagram, 
we find that most of the brown dwarfs are less than 1 Myr, but find a number 
of low mass stars with inferred ages greater than 3~Myr.  The discrepancy in 
the stellar and substellar ages is due to our selection of only low luminosity 
sources; however, the presence of such objects implies the presence of an 
age spread in the OMC 2/3 region.  We discuss possible reasons for this 
apparent age spread.

\end{abstract}

\keywords{infrared: stars -- open clusters and associations: individual (\object{Orion Molecular Clouds 2 and 3}, \objectname{OMC 2/3}) -- stars: formation -- 
stars: imaging -- stars: low-mass, brown dwarfs -- techniques: spectroscopic}

\section{Introduction}

The highly active star forming region known as the Orion Molecular Cloud 2/3 
region (hereafter, OMC 2/3) is located in a molecular filament \citep{b87} 
in the northernmost part of the Orion A molecular cloud at a distance of 450 
pc \citep{gs89}.  This region contains one of the richest concentrations of 
protostars and prestellar cores known.  Thirty-four submillimeter sources have 
been detected in OMC 2/3 at 1300 $\mu$m and 350 $\mu$m, evidence that it is 
a site of vigorous star formation activity \citep{chini97,lis98}.  All six 
condensations in OMC 3, the northernmost part of OMC 2/3, fit the definition 
of Class 0 object, L$_{bol}$/L$_{submm}$ $<$ 200, and thus are in the earliest 
stages of stellar evolution \citep{chini97}.  In addition, at least 80 knots 
of H$_2$ $v = 1 - 0$ $S(1)$ emission, signifying jets and outflows, have been 
detected in OMC 2/3, confirming the region's extreme youth \citep{ybd97}.

Although OMC 2/3 has such a high density of star-forming cores, as shown by 
the richness of submillimeter sources, the population of pre-main sequence 
(PMS) stars and young brown dwarfs has not been studied as extensively or to 
similar depth in the infrared as the Orion Nebula Cluster (ONC), which is 
directly south of OMC 2/3 \citep[e.g.][]{shc04,lrah01}.  As part of a study 
to survey the PMS star population in OMC 2/3, we present the results of a 
survey of young brown dwarfs.  With the existence of brown dwarfs firmly
established, attention has shifted towards understanding the formation 
mechanism.  \citet{rc01} proposed that brown dwarfs are formed by the ejection 
of accreting ``stellar embryos'' from dynamically unstable systems of 
protostars, resulting in the premature termination of accretion and substellar 
masses.  Our study of brown dwarfs in OMC 2/3 was initially motivated by this 
proposal; the detection of a large number of brown dwarfs around Class 0 
sources would be evidence for ejection at slow velocities.  \citet{rc01} 
suggest that given the amount of star formation activity, this would be the 
best place to find both very young brown dwarfs and potentially proto-brown 
dwarfs, and hence test models of their formation.  A more detailed discussion 
of brown dwarf formation in OMC 2/3 will be discussed in a subsequent paper on 
{\it Spitzer} colors of the confirmed young brown dwarfs presented in this 
paper.

In order to obtain a complete census of the young brown dwarfs in OMC 2/3 (and 
eventually determine whether they are likely to have circumstellar disks), 
near-infrared observations are critical for identification.  We performed 
near-infrared observations that are two to three magnitudes deeper than the 
Two Micron All Sky Survey \citep[2MASS;][]{2mass06} in order to identify the 
most deeply embedded substellar objects in OMC 2/3, down to 25 \mjup, based on 
1~Myr isochrones \citep{bcah98}.  Before 2MASS, near-infrared observations of 
OMC 2/3 reached $K-$band magnitudes of 12.20 magnitudes \citep{jgjhg90}; as a 
comparison, for our near-infrared observations we reach a magnitude limit of 
$K =$ 17.9.

Near-infrared and visible wavelength imaging and spectroscopy provide the 
information necessary for the selection and confirmation of substellar 
objects in star-forming regions.  Young brown dwarfs, with ages much less 
than 10~Myr, are much more luminous and thus easier to detect than field 
brown dwarfs (typically older than 1~Gyr) at distances greater than 100 pc.  
However, the near-infrared colors of young brown dwarfs are similar to those 
of very low mass stars making it difficult to identify brown dwarfs from 
near-infrared photometry alone.  Therefore techniques developed by \citet{l00} 
are applied to select brown dwarf candidates in OMC 2/3 from combined, deep 
near-infrared and visible wavelength photometry.  Once candidate brown dwarfs 
are photometrically selected, far-red (0.6$-$1.0 $\mu$m) and near-infrared 
(0.8$-$2.5 $\mu$m) spectroscopic observations are used to confirm the 
candidates as {\it bona fide} brown dwarfs.

We present a multiwavelength analysis of the OMC 2/3 region to identify 
the population of young brown dwarfs.  In Section \ref{sec:observations}, we 
introduce the ground-based imaging of the OMC 2/3 region.  Next, in Section 
\ref{sec:selection}, we describe the techniques used for selecting candidate 
young brown dwarfs for spectroscopy using the multiwavelength photometry.  
Then, we describe the spectroscopic observations (Section \ref{sec:specobs}) 
and the spectral type classification methods used as well as the spectral 
types determined from the spectra (Section \ref{sec:spectral_typing}).  
Finally, we discuss the age of OMC 2/3 in Section \ref{sec:omc23age}.

\section{Observations: Imaging \label{sec:observations}}

\subsection{Near-Infrared Imaging \label{sec:sqiidimaging}}

\begin{figure}
\plotone{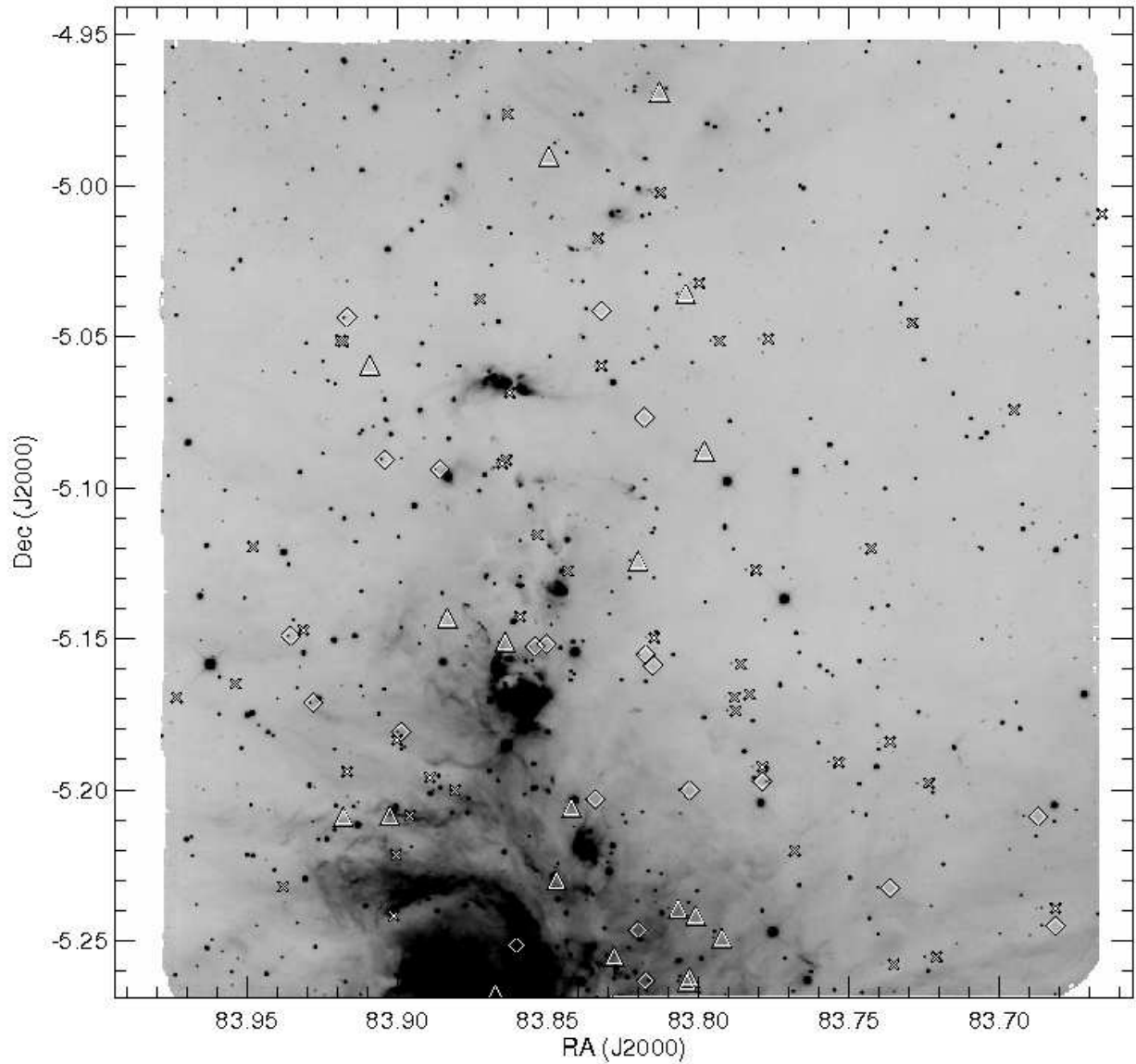}
\caption[Spatial Distribution: SQIID $K-$band image of OMC 2/3]{An 
inverse-scale SQIID image of OMC 2/3 in the $K-$band.  The distribution of the 
40 low new low mass members is shown: the 19 sources with spectral types of 
M6.5$-$M9 from Table \ref{table:browndwarfs} (triangles), and the 21 sources 
with spectral types of M4$-$M6 from Table \ref{table:lowmassmemberspectra} 
(diamonds).  For reference, we also show the sources from Table 
\ref{table:fieldspectra} (crosses) which are field stars or whose spectral 
types we were unable to determine.}
\label{fig:image_dist}
\end{figure}

The near-infrared observations of OMC 2/3 were obtained with SQIID 
\citep{ellis92}, the Simultaneous Quad Infrared Imaging Device on the 
2.1-meter telescope at Kitt Peak National Observatory during the period of 
2000 December 11$-$13.  The conditions were non-photometric, with thin clouds, 
and the seeing was around 2$^{\prime\prime}$ (closer to 1.8$^{\prime\prime}$ 
in $K-$band).  SQIID utilizes a 1024 $\times$ 1024 InSb detector array with 
independent 512 $\times$ 512 quadrants that are filtered at Barr J (1.27 
\micron), Barr H (1.67 \micron), Barr K (2.22 \micron) and Barr PAH (3.30 
\micron).  As seen in Figure \ref{fig:image_dist}, our SQIID image mosaic 
toward OMC 2/3 totals a 20$^{\prime}$ $\times$ 20$^{\prime}$ field of view 
centered slightly north of OMC 2, $\alpha$ = 05$^h$ 35$^m$ 22.8$^s$, $\delta$ 
= $-05^{\circ}$ 6$^{\prime}$ 37.1$^{\prime\prime}$ (J2000), at $J$, $H$ and 
$K$ (no useful data were obtained at 3.30 $\mu$m with SQIID).  These 
wide-field mosaics were obtained by mapping the region in a 4 $\times$ 4 grid 
with approximately 90\arcsec~ of overlap between images in adjacent grid 
positions.  Each image in the grid has 512 $\times$ 512 pixels; with a 
platescale of 0.69\arcsec/pixel, the total size of the final mosaic is 
20$^{\prime}$ $\times$ 20$^{\prime}$.  The 4 $\times$ 4 map was repeated 
seven times with slightly offset positions each time; the typical total 
integration time for each image is 7 minutes.

Customized Interactive Data Language (IDL) programs were implemented to 
perform the image reduction including linearity, flat fielding, background 
subtraction, the elimination of bad pixels, and distortion correction, as well 
as creation of individual mosaics and aperture photometry.  Calibration of the 
SQIID observations was done by comparison with 2MASS photometry; thin clouds 
were present during the observations, and thus the data were not calibrated 
with standard stars.  To properly compare the 2MASS stellar magnitudes with 
the magnitudes of the sources in our SQIID images, 15$-$30 stars were chosen 
for calibration in each of the sixteen fields.  Stars which were determined by 
\citet*{chs01} to show variability were not among those included for 
calibration.  In addition, sources brighter than the saturation limits of 
$J=$ 10.31, $H=$ 10.06 and $K=$ 9.45 as well as sources fainter than the 2MASS 
completeness limits were not included.  Each 512 $\times$ 512 pixel frame 
was calibrated individually by finding the offset between the instrumental 
magnitudes derived in that frame and the magnitudes from the 2MASS Point 
Source Catalog for each selected source.  The rms scatter of the residuals 
between the 2MASS and SQIID data sets ranges from 0.05 to 0.08 magnitudes; 
this is on the order of the typical uncertainty in the instrumental magnitude 
for SQIID (for which the maximum allowed was 0.15 magnitudes).  After the 
calibrated frames were combined into mosaics, we compared the photometry from 
our calibrated SQIID mosaics with 2MASS to look for systematic offsets between 
the two systems (mainly due to the difference in the K bandpasses) and found 
the following relations:

\begin{equation}
J_{2MASS} - J_{SQIID} = -0.0039(\pm 0.0026) (J - K)_{SQIID} - 0.014(\pm 0.004)
\end{equation}
\begin{equation}
H_{2MASS} - H_{SQIID} = -0.026(\pm 0.002) (J - K)_{SQIID} + 0.020(\pm 0.004)
\end{equation}
\begin{equation}
K_{2MASS} - K_{SQIID} = 0.017(\pm 0.001) (J - K)_{SQIID} - 0.023(\pm 0.002)
\end{equation}

\noindent
where $K_{2MASS}$ is a $K_S$ filter.  Over 350 stars in each band were 
used to calculate these relationships; all have photometric uncertainties 
less than 0.10 magnitudes.

Once the images were calibrated, IDL routines including PhotVis, a 
graphical interface for the IDL aperture photometry routine \citep{rguter04}, 
were used to perform aperture photometry on the final image mosaics.  A 
synthetic aperture radius of 3 pixels was used with a small sky annulus of 
2 pixels in width (identical to that used for calibration).  The final source 
list contained 823 sources that appeared in all three bands with photometric 
uncertainties less than 0.15 magnitudes; only sources with \jhk photometric 
uncertainties less than 0.15 magnitudes were used for analysis.  Note that 
the photometric uncertainties tabulated throughout this paper do not take into 
account the uncertainties in the absolute calibration discussed in the 
previous paragraph.  The tabulated uncertainties are calculated (by PhotVis) 
from the expected shot noise and read noise and the measured RMS variations in 
the background annulus.  

The limiting magnitudes reached with the detection criteria were $J =$ 19.4, 
$H =$ 18.3 and $K =$ 17.9.  We derived 90\% completeness limits by adding 
synthetic stars to our images for a range of magnitudes using a Gaussian PSF 
with the same FWHM as that of our images and performing the same point source 
extraction.  Measurements were rejected if they had photometric uncertainties 
greater than 0.15 magnitudes and if the recovered magnitude was more than 
3$\sigma$ brighter than the input magnitude, indicating that a star was likely 
coincident with a synthetic star.  Table \ref{table:complete_summary} lists 
the magnitude at which 90\% of the stars were recovered for the $J-$, $H-$ and 
$K-$bands.  For comparison, published completeness limits for 2MASS are 
$J=15.8$, $H=15.1$ and $K_{s}=14.3$ magnitudes \citep[where source counts are 
$>$ 99\% complete at the S/N=10 limits;][]{2mass06}.

\placefigure{fig:image_dist}

\subsection{\izcap Imaging \label{sec:4shooterimaging}}

The visible wavelength \iz images were obtained under photometric conditions 
with $\sim$2.5$^{\prime\prime}$ seeing, 2002 December 7 using the 4-Shooter 
camera on the 1.2-meter telescope at Fred Lawrence Whipple Observatory 
(FLWO).  The 4-Shooter camera has four 2048$\times$2048 arrays, arranged in a 
2$\times$2 grid with 45\arcsec~ separation between chips.  The images were 
binned 2$\times$2 during readout, resulting in a pixel scale of 
0.67\arcsec/pixel.  The final 20$^{\prime}$ $\times$ 20$^{\prime}$ field was 
centered to match the SQIID \jhk images.  OMC 2/3 was imaged at Harris I (8100 
\AA) and $z^{\prime}$ (9100 \AA), twice at each band with an integration time 
of 20 minutes each, for a total of 40 minutes at each wavelength.  Similar 
steps were taken to reduce the \iz images as for the SQIID \jhk images, 
including bias subtraction, flat fielding, the elimination of bad pixels, and 
distortion correction. 

The standard stars used to calibrate the 4-Shooter images of OMC 2/3 
are from the \citet{landolt92} SA98-670 and SA98-618 fields.  The 
$I-$band data were calibrated from standards in the Cousins I system without 
including a color dependence term because the $I-$band filter at FLWO is 
similar.  Since $z^{\prime}$ magnitudes are not tabulated in 
\citet{landolt92}, the $z^{\prime}$ magnitudes were estimated from the 
$I-$band magnitudes of standard stars with neutral colors.  The stars in the 
\citet{landolt92} fields with $R-I$ colors less than 0.4 were used, and 
$I-z^{\prime}$ = 0 assumed.  Using this calibration for our $z^{\prime}$ 
magnitudes is sufficient because these data are only used to identify young 
low mass candidates from background sources.

Each standard field was imaged twice over the course of the night so that an 
airmass correction could be accounted for in the calibration.  Aperture 
photometry was obtained using a radius of 6 pixels with a sky annulus of 4 
pixels in width for both the standard stars used for calibration and for the 
OMC 2/3 sources.  The rms scatter of the residuals for the \citet{landolt92} 
standard stars imaged with the 4-Shooter is less than 0.05 magnitudes; this 
calibration uncertainty is on the order of the typical intrinsic photometric 
uncertainties for the 4-Shooter observations.  Only sources with photometric 
uncertainties less than 0.15 magnitudes were used in our analysis (as with the 
\jhk data, calibration uncertainties are not included in the tabulated 
uncertainties).  Completeness limits for \iz were analyzed in the same way as 
for the \jhk data (see Section \ref{sec:sqiidimaging}) and are also listed in 
Table \ref{table:complete_summary}.  The completeness limits for the visible 
wavelength photometry were difficult to assess because of nebulosity from the 
ONC in the southern part of the image.  Therefore the 90\% completeness limits 
quoted in Table \ref{table:complete_summary} are for the region north of a 
declination of $-5.15$; if we include the entire image in our analysis, the 
magnitudes quoted in Table \ref{table:complete_summary} are 80\% complete.  

\section{Photometric Selection of Brown Dwarf Candidates \label{sec:selection}}

Candidate young brown dwarfs in OMC 2/3 are selected from the merged 
SQIID and 4-Shooter \ijhk photometry.  Brown dwarfs are identified by two 
criteria: their low luminosity and, when they are detected in the $I-$band, 
their red colors.  These candidates still require confirmation with 
spectroscopic observations, though, to eliminate contamination from faint 
background stars and to distinguish between older low mass stars both in OMC 
2/3 and in the foreground.  The candidates are selected for follow-up 
spectroscopy in the far-red, and in cases where they are too faint to detect 
in the $I-$band, in the near-infrared.

\begin{figure}
\plotone{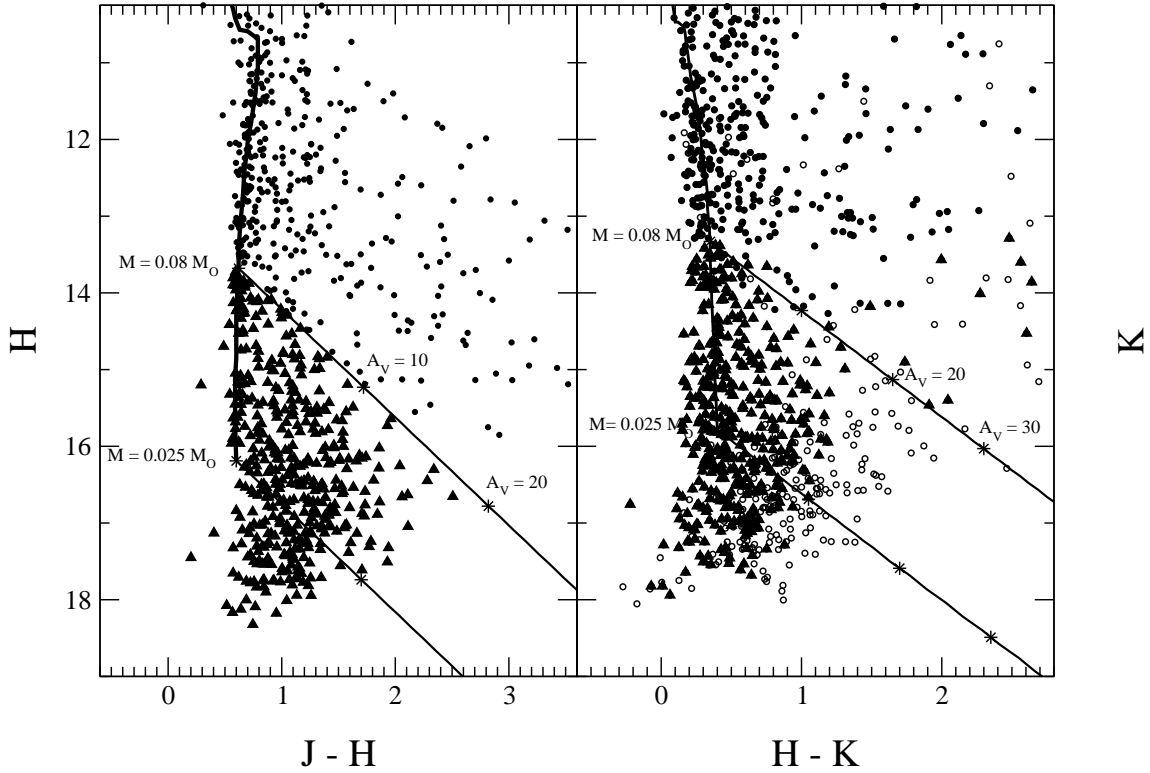}
\caption[Brown Dwarf Candidates: Near-Infrared Color-Magnitude Diagrams]{These 
two color-magnitude diagrams show all sources detected at \jhk in the SQIID 
image with photometric uncertainties less than 0.15 magnitudes (filled 
circles).  The $H$ versus $J-H$ CMD on the left identifies, as filled 
triangles, the brown dwarf candidates selected based on their location near or 
below the HBL.  Over-plotted is the 1~Myr \citet{bcah98} evolutionary 
isochrone, assuming a distance to OMC 2/3 of 450 pc.  In addition, reddening 
vectors \citep{cfpe81} are plotted for a 0.08 \msun~source with an age of 
1~Myr (upper diagonal line) and a 25 \mjup~brown dwarf with an age of 1 Myr 
(lower diagonal line).  The plot on the right shows, with the same symbols, 
the brown dwarf candidates on a $K$ versus $H-K$ CMD.  Many of the red sources 
near the HBL on the $H$ versus $J-H$ CMD are above the HBL on the $K$ versus 
$H-K$ CMD; these sources may have strong infrared excesses.  The open circles 
represent sources detected in the $H$ and $K-$bands (and not in $J$).}
\label{fig:double2}
\end{figure}

\placefigure{fig:double2}

To search for brown dwarfs on the basis of their low luminosities, candidates 
are initially selected from their positions on an $H$ versus $J-H$ 
color-magnitude diagram (CMD), as shown in Figure \ref{fig:double2}. The 
models of \citet{chabrier97} show that the hydrogen burning limit (HBL) for 
solar metallicity objects is at 0.075 M$_{\odot}$.  However, given the 
uncertainty in the ages of the sources, we consider a ``soft boundary'' of 
0.080 M$_{\odot}$ to be used to guide the selection of objects and not as the 
precise location of the HBL.  To find the corresponding luminosity, an age 
and distance must be adopted and we adopt a distance of 450 pc and an age of 
1~Myr.  Several recent studies of the distance to Orion place the distance 
closer, consistent with a value of 420 pc \citep{h07,jef07,mrfb07,sand07}.  
Since this result was published after our analysis, we use 450 pc throughout 
this paper.  Uncertainties that arise from this are discussed in Section 
\ref{sec:lbol}.  Adopting an age of 1~Myr is motivated by both the 
concentration of young protostellar objects in this region and the age of the 
neighboring ONC ($<$ 1$-$2 Myr; \citet{hillenbrand97}).  Using the 1~Myr 
evolutionary tracks of \citet{bcah98} and the spectral type versus effective 
temperature scale from \citet{l03}, the spectral type corresponding to a mass 
of 0.080 M$_{\odot}$ is approximately M6$-$M6.5; in this paper, we use M6.5 as 
the fiducial boundary between stellar and substellar objects.

\begin{figure}
\plotone{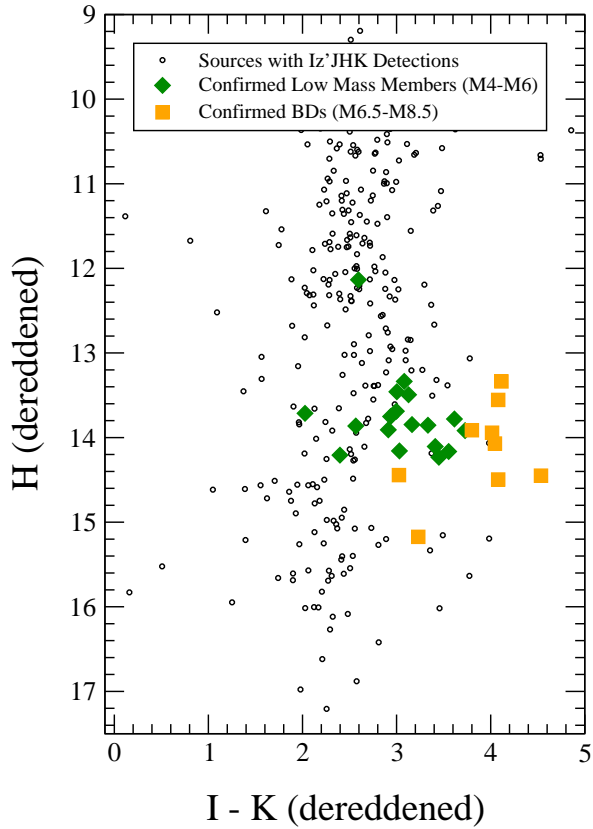}
\caption[Dereddened CMD for OMC 2/3]{Dereddened color-magnitude diagram 
for OMC 2/3 assuming an intrinsic $J-H$ color of a late M star for each 
source (see Appendix \ref{sec:lbol_teff}).  The positions of known brown 
dwarfs define the region of the color-magnitude plane populated by young brown 
dwarfs at the distance of Orion.  We find that the spectroscopically 
verified brown dwarfs in OMC 2/3 (filled squares; colored orange in electronic 
edition) occupy a very limited region of the color-magnitude plane.  See the 
electronic edition of the Journal for a color version of this figure.}
\label{fig:dereddened}
\end{figure}

\placefigure{fig:dereddened}

\begin{figure}
\plotone{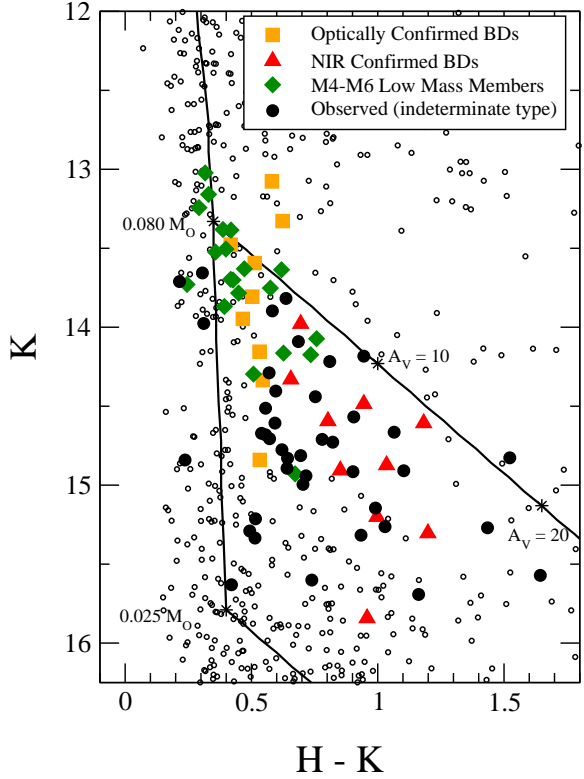}
\caption[Near-Infrared Selection of Brown Dwarfs]{$K$ vs. \hk~ 
color-magnitude diagram showing all 1017 sources detected at $H$ and $K$ with 
uncertainties less than 0.15 magnitudes in OMC 2/3 (open black circles).  
Over-plotted is the \citet{bcah98} evolutionary isochrone for a 1 Myr-old 
pre-main sequence star, assuming a distance to OMC 2/3 of 450 pc.  Reddening 
vectors are plotted for: 1) a 0.08 \msun~star (upper diagonal line) and 2) a 
25 M$_J$ brown dwarf with an age of 1 Myr (lower diagonal line).  Brown dwarfs 
confirmed by far-red and near-infrared spectroscopy are denoted by filled 
squares (colored orange in electronic edition) and triangles (colored red in 
electronic edition), respectively.  Sources denoted by filled diamonds 
(colored green in electronic edition) are the 21 candidates confirmed by 
spectroscopy to be M4$-$M6 low mass members.  Sources denoted by filled 
circles were observed with SpeX during the December 2003, January 2005 and 
2006 runs; they exhibit featureless spectra, which we argue makes them likely 
background sources with spectral types earlier than M4.  Note that some brown 
dwarfs were found with magnitudes above the HBL.  These may be due to ages 
younger than the assumed 1~Myr age and/or the presence of infrared excess.  
See the electronic edition of the Journal for a color version of this figure.}
\label{fig:irtfpropfigs2003}
\end{figure}

\placefigure{fig:irtfpropfigs2003}

At the faint magnitudes typical of young brown dwarfs, contamination by 
background stars is considerable.  Studies of the Taurus and IC 348 young 
clusters \citep{l00,l99} have shown that members of a young population can be 
separated from background stars using visible and near-infrared wavelength 
CMDs (e.g. $H$ versus $I-K$).  As \citet{l00} showed with the Taurus region, 
background stars follow a different locus than cluster members.  Therefore the 
\citet{l00} techniques of using combined, deep near-infrared and visible 
wavelength photometry to select brown dwarf candidates for spectral 
observations are used here for OMC 2/3.  

For the candidate selection, we first placed all the sources detected at 
\ijhk that fell below the 0.080 M$_{\odot}$ reddening vector in the $H$ versus 
$J-H$ CMD (Figure \ref{fig:double2}) on a $I-K$ versus $J-H$ color-color 
diagram and a dereddened, $H$ versus $I-K$ CMD.  Figure \ref{fig:dereddened} 
shows how the spectroscopically confirmed brown dwarfs in OMC 2/3 are located 
in a very limited region of a dereddened, $H$ versus $I-K$ CMD, as was also 
seen in Taurus \citep{l00}.  We note that there are a few sources located in 
this region of the CMD with colors and magnitudes similar to the brown dwarfs 
which we have not obtained spectra of because they were too faint.  For the 
initial candidate list, we selected sources with dereddened $H-$band 
magnitudes greater than 13 and with dereddened $I-K$ colors greater than or 
equal to 3.  Although a $I-K$ versus $J-H$ color-color diagram and $I$ versus 
$I-z^{\prime}$ CMD are not shown here, they were also useful for candidate 
selection when comparing with reddening vectors for sources less massive than 
0.080 M$_{\odot}$.  The dereddened $I-K$ color is used as a proxy for 
temperature for sources with $I-$band detections.  The dereddening is 
performed using the extinction law of \citet{cfpe81} and adopting a typical 
$(J-H)_{0}$ color of 0.6 (see Appendix \ref{sec:lbol_teff} for a complete 
description).  By identifying sources with $I-K$ colors consistent with young 
brown dwarfs ($I-K \geq$ 3), as seen in Taurus, we reduce contamination both 
from faint background stars and from older low mass stars in OMC 2/3 and the 
foreground.

\begin{figure}
\plotone{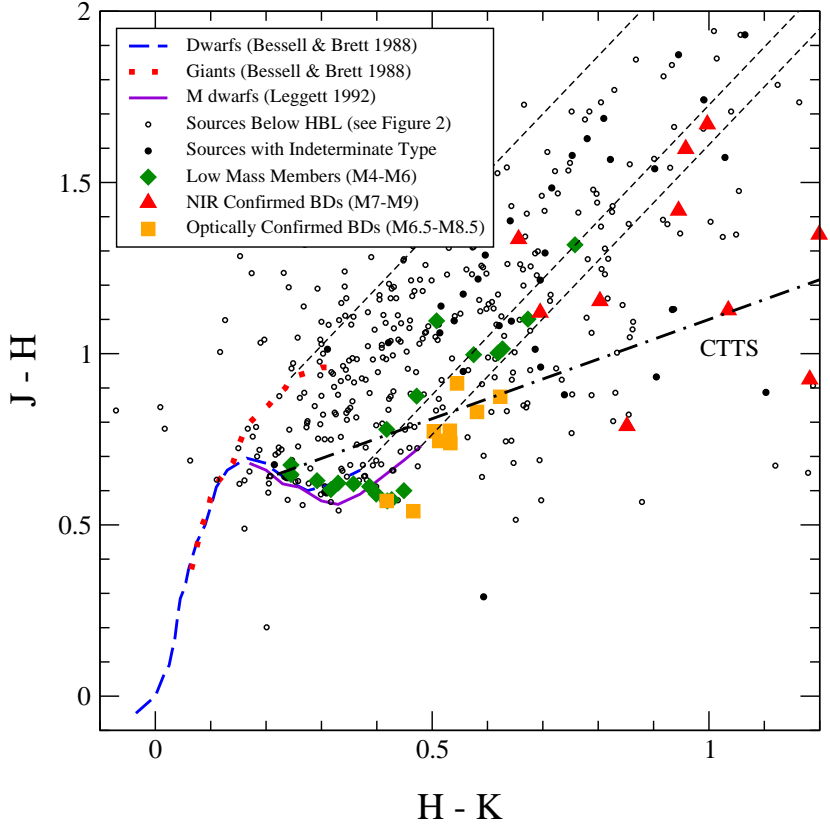}
\caption[$JHK-$Band Color-Color Diagram: OMC 2/3]{The above $J-H$ versus $H-K$ 
color-color diagram for OMC 2/3 shows SQIID photometry for the brown dwarf 
candidates: the sources with color magnitude errors less than 0.15 
magnitudes that appear below the 0.08 M$_{\odot}$ reddening line in the $H$ 
versus $J-H$ CMD (see Figure \ref{fig:double2}).  Those sources that are 
either field stars or whose spectral types could not be determined (see Table 
\ref{table:fieldspectra}) are shown as filled circles.  The sources 
spectroscopically confirmed as M4$-$M9 members are compared with, from 
\citet{bb88}, the colors of giants (dotted locus colored red in 
electronic edition) and dwarfs (dashed locus colored blue in electronic 
edition), as well as M dwarfs from  \citet[][solid line colored purple in 
electronic edition]{leggett92}.  (Note that these colors have not been 
transformed into the SQIID photometric system.)  Small dashed lines indicate 
reddening vectors \citep{cfpe81} for (from left) M5 giants, M6 dwarfs, and M9 
sources (typical colors for M9 stars from \citet{kirk00}).  The dot-dashed 
line is the classical T Tauri star locus from \citet*{mch97} in Taurus.  See 
the electronic edition of the Journal for a color version of this figure.}
\label{fig:nir_cc}
\end{figure}

For highly extinguished sources (i.e. sources that did not have $I-$band, and 
in some cases, $J-$band detections), we rely on their dereddened luminosity 
and colors in the near-infrared alone, which makes it difficult to distinguish 
cool, substellar members from background stars: giants as well as older, 
higher temperature dwarfs along the line of sight.  The candidate brown dwarfs 
selected for near-infrared spectroscopy were chosen because they met several 
criteria.  First, their $K-$band magnitudes range from 13 to 16 magnitudes 
(where K = 16 is the magnitude limit reached with SpeX) and they fall below 
the 0.080 M$_{\odot}$ reddening vector on the $J$ versus \jh~ CMD 
and/or $K$ versus \hk~ CMD (see Figure \ref{fig:irtfpropfigs2003}).  In the 
CMD, brown dwarfs confirmed from far-red spectroscopy are denoted by filled 
squares (note that chronologically, the far-red spectra were obtained earlier, 
so brown dwarf status was already confirmed for some sources when the 
near-infrared candidates were selected and observed).  Also, many of the 
candidate brown dwarfs selected for near-infrared spectroscopy were chosen 
because they exhibit $A_V > 7$.  This criterion limits overlap with the 
optically visible sample and reduces contamination from foreground stars.  
Additionally, the sources with high S/N $J-$band magnitudes were placed on a 
near-infrared color-color diagram as seen in Figure \ref{fig:nir_cc} to reject 
sources with colors similar to those of giants 
\citep[][red dotted line in Figure \ref{fig:nir_cc}]{bb88}.

\placefigure{fig:nir_cc} 

\section{Spectroscopic Observations \label{sec:specobs}}

Spectra were obtained for a total of 88 brown dwarf candidates identified in 
OMC 2/3.  This includes 32 candidates obtained with the MMT and Keck 
telescopes in the far-red spectral regions, and 56 highly extinguished 
candidates observed in the near-infrared regime with the Infrared Telescope 
Facility (IRTF).

\subsection{Far-Red Spectroscopy \label{sec:farredspectra}}

Far-red (0.6$-$1.0 $\mu$m) spectral observations were obtained for a total of 
32 brown dwarf candidates in OMC 2/3 on four different occasions.  Spectra of 
seven brown dwarf candidates were obtained 2002 November 5 with the Low 
Resolution Imaging Spectrograph \citep{oke95} on the 10-meter Keck I telescope 
from about 6400 to 7700\AA~ using the 1200 lines/mm grating; spectra were 
obtained with about a 2 \AA~resolution.  Spectra of twelve brown dwarf 
candidates were obtained on 2002 November 11 and 2002 December 15 using the 
Blue Channel Spectrograph on the MMT Observatory 6.5-meter telescope, centered 
at 7600 \AA~using the 600 lines/mm grating.  Lastly, spectra of thirteen brown 
dwarf candidates were obtained on 2003 January  26$-$27 with the Red Channel 
Spectrograph at the MMT, also centered at 7600 \AA~using the 270 lines/mm 
grating.  

In addition to the spectra of the candidates, arc (HeNeAr) lamp spectra were 
obtained for wavelength calibration and continuum ``bright'' lamps were 
observed for flat fielding.  All of the far-red spectra are reduced 
similarly.  First a bias frame is subtracted from the source frames, then they 
are flat fielded using the spectra of the bright lamps.  Finally the source 
spectrum is extracted and wavelength calibrated using the HeNeAr lamp 
spectrum.  

\subsection{Near-Infrared Spectroscopy \label{sec:nirspectra}}

\subsubsection{SpeX Data \label{sec:spexspectra}}

For the highly extinguished OMC 2/3 brown dwarf candidates, low resolution, 
near-infrared spectroscopy was obtained with the SpeX spectrograph at the 
Infrared Telescope Facility (IRTF) on Mauna Kea \citep{rayner03}.  A total of 
56 candidates were observed with SpeX, where four of them overlap with the 
aforementioned candidate spectra obtained in the far-red.  Observations with 
SpeX were made in single-prism mode ($\lambda$/$\Delta\lambda$ $\sim$ 250) on 
the nights of 2003 December 21$-$23, 2005 January 5, and 2006 January 3$-$5.  
Near-infrared spectroscopy is the only reliable means for obtaining an 
accurate census of the more heavily embedded (and therefore reddened) brown 
dwarfs in OMC 2/3 and the wavelength coverage of SpeX in the single-prism mode 
(0.8$-$2.5 $\mu$m) combined with the high, dry site of Mauna Kea is ideal for 
accurately measuring water absorption in young brown dwarfs.  This is 
important because the spectral types of our OMC 2/3 brown dwarf candidates 
observed with SpeX are established from the depth of the water vapor 
absorption bands \citep*{wgm99}, as will be described in Section 
\ref{sec:spectral_typing}. 

In addition to the 56 OMC 2/3 candidates that have been observed with SpeX to 
date, 23 optically classified standards have also been observed.  These 
standards are young brown dwarfs and low mass stars (in OMC 2/3 and Taurus), 
field dwarfs, and giants \citep{briceno02}.  These comparison spectra were 
obtained in order to quantify the shape of the water absorption features 
indicative of late M-type stars, and thus provide a set of spectral 
calibration templates.   

The SpeX spectra obtained 2003 December 21$-$23 and 2005 January 5 were 
reduced using the Spextool package version 3.2 \citep*{cushing04} and those 
obtained 2006 January 3$-$5 were reduced using version 3.4.  Several spectra 
from the 2003 observing run were re-reduced with version 3.4 for comparison; 
no difference was found.  The Spextool package performs the basic reduction 
steps described in the previous section for the reduction of the far-red 
spectra.  Observations of the nearby A0V star, HD 37887, were obtained between 
each source to facilitate the removal of telluric features.  The Spextool 
package uses a high resolution model of Vega to remove the hydrogen absorption 
features in the stellar spectrum, as described by \citet*{vacca2003}.

\subsubsection{CorMASS Data}

Six near-infrared spectra were obtained using CorMASS (where two of them 
overlap with candidate spectra obtained in the far-red).  CorMASS, the Cornell 
Massachusetts Slit Spectrograph \citep{wilson01}, is a cross-dispersed 
spectrograph with a $\lambda$/$\Delta\lambda$ of approximately 300 and 
wavelength coverage of 0.75$-$2.5 $\mu$m, similar to SpeX.  One object was 
observed 2005 May 1 at the 6.5-meter Magellan (Clay) Telescope at Las Campanas 
Observatory.  Three objects were observed on the dates of 2004 October 23 and 
2005 January 29 at the 1.8-m Vatican Advanced Technology Telescope at Mount 
Graham International Observatory.  Two objects were observed on the dates of 
2005 October 16 and 2006 January 18 at the Apache Point Observatory 
(APO) 3.5-meter telescope.

Data reduction was performed using Cormasstool, a modification of Spextool 
\citep*{cushing04}, again including the basic reduction steps described in the 
previous sections.  The standard A0V star (HD 37887) was used to remove 
the telluric features.

\section{Spectral Classification \label{sec:spectral_typing}}

\subsection{Far-Red Spectra}

\subsubsection{Molecular Features \label{sec:molecular}}

\begin{figure}
\plotone{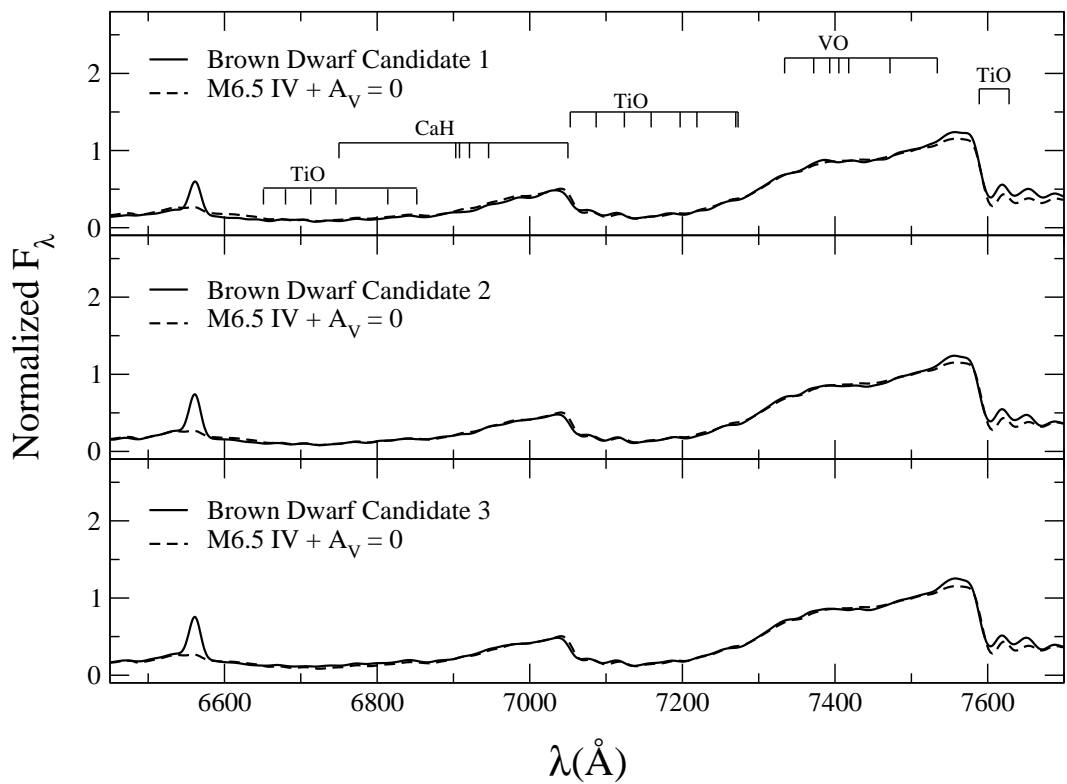}
\caption[Keck LRIS Spectra]{Candidate brown dwarf spectra obtained with LRIS 
on Keck I.  The over-plotted dotted lines show the reference spectra used to 
classify the candidates.  All three of these candidates look very 
similar: each has an M6.5 spectral type with little or no reddening and all 
exhibit H$\alpha$ (at 6563 \AA).  In addition, they all match the average 
spectrum of a M6.5 dwarf and giant: indicating a young source.  The TiO, 
CaH and VO molecular bands in this wavelength range used for spectral 
classification are indicated.  Each of the spectra have been smoothed to a 
resolution of 18 \AA~ and normalized at 7500 \AA.}
\label{fig:staufspectra}
\end{figure}

\begin{figure}
\epsscale{.80}
\plotone{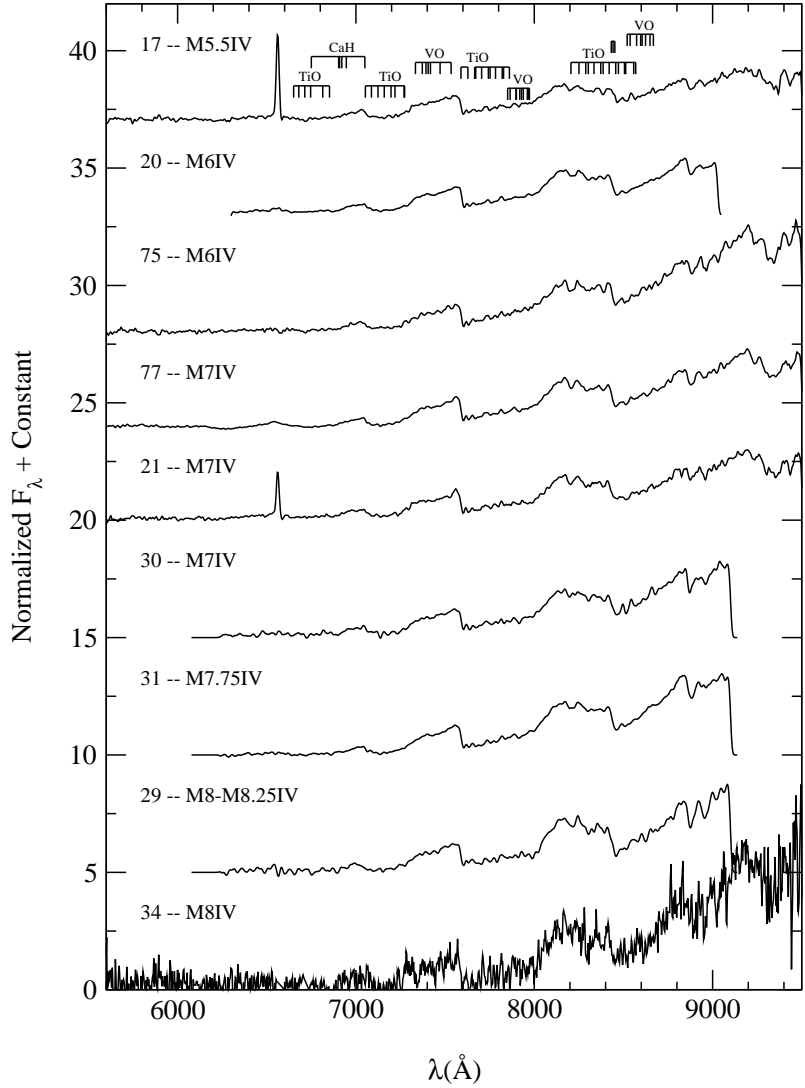}
\caption[MMT Far-Red Spectra]{Low resolution spectra from the MMT with the 
Red (5500$-$9800 \AA) and Blue (6200$-$9000 \AA) Channel Spectrographs of nine 
new low mass members in OMC 2/3.   The TiO, CaH, and VO molecular bands used 
for spectral classification are indicated.  Strong H$\alpha$ emission can be 
seen in two of the sources: 17 and 21.  The spectra have been smoothed to a 
resolution of 18 \AA~ and normalized at 7500 \AA.}
\label{fig:mmtspectra}
\end{figure}

\begin{figure}
\epsscale{.80}
\plotone{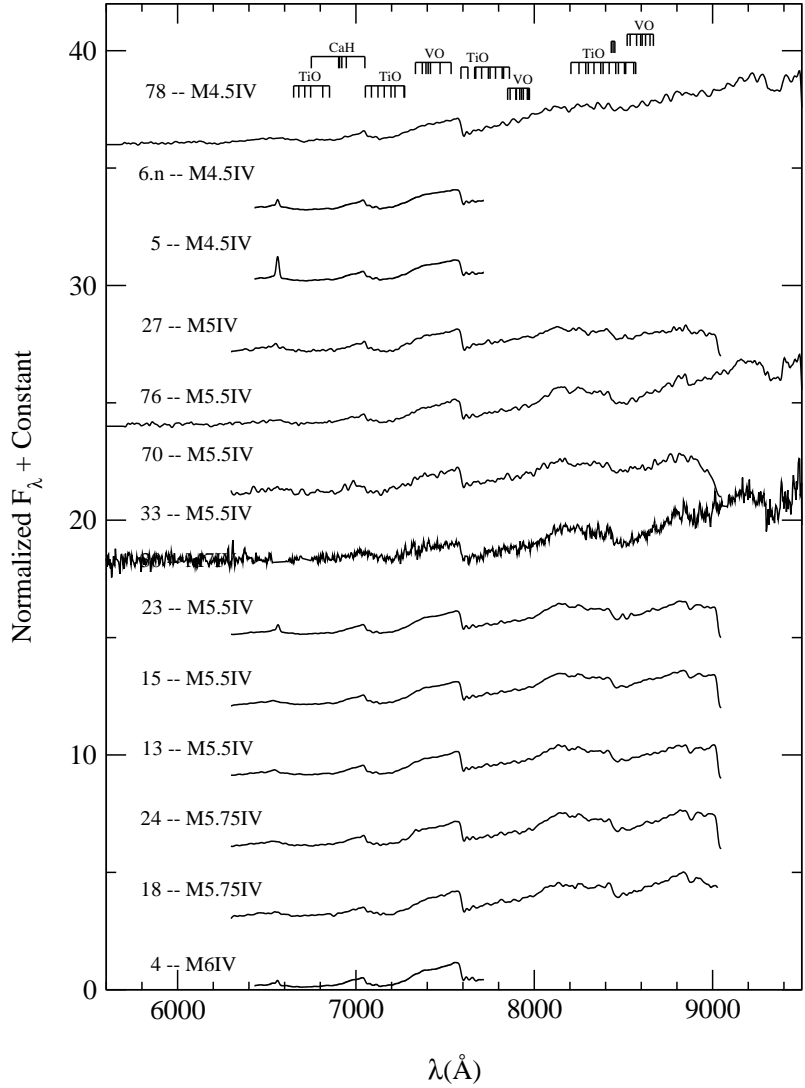}
\caption[Far-Red Spectra]{Low resolution spectra obtained with LRIS 
(6400$-$7700 \AA) on Keck I as well as with the Red (5500$-$9800 \AA) and Blue 
(6200$-$9000 \AA) Channel Spectrographs on the MMT of thirteen new low mass 
members in OMC 2/3.   The TiO, CaH, and VO molecular bands used for spectral 
classification are indicated.  The spectra have been smoothed to a resolution 
of 18 \AA~ and normalized at 7500 \AA.  }
\label{fig:mmtspectra2}
\end{figure}

Spectral types are determined by the relative strengths of specific atomic 
and molecular features.  The wavelength range of the far-red spectra we 
obtained, 0.6$-$1.0 $\mu$m, includes several distinctive absorption features 
commonly seen in low mass stars and brown dwarfs: titanium oxide (TiO) and 
vanadium oxide (VO) as well as calcium hydride (CaH).  In general, each of the 
candidates was given a spectral type based on the depth of the TiO and VO 
molecular bands (Figures \ref{fig:staufspectra}, \ref{fig:mmtspectra}, and 
\ref{fig:mmtspectra2}).  Note how the TiO absorption features around 7600 and 
8400 \AA~are deeper at later M spectral types (Figures \ref{fig:mmtspectra} 
and \ref{fig:mmtspectra2}).

To assign spectral types and determine whether a given star is a pre-main 
sequence member, a background giant, or a foreground dwarf, the spectrum was 
compared against spectra of standards with known spectral types.  These 
standards included spectra of dwarfs, giants, and the average of a giant and 
dwarf at a given M spectral type.  As demonstrated in \citet{l99}, the depths 
of the TiO, VO, and CaH molecular bands seen in the pre-main sequence low mass 
stars are fit well by the average of dwarf and giant spectra.  Young stars and 
brown dwarfs have lower surface gravities than more evolved field dwarfs but 
do not have surface gravities as low as giants.  For objects with spectral 
types later than M5, averaging a giant and a dwarf result in a good fit 
because gravity sensitive features seen in field dwarfs are weaker in the 
spectra of the young stars and brown dwarfs \citep{l99}.  Not all features are 
dependent on surface gravity.  Absorption in VO (at 7900 and 8500 \AA), a 
feature used for spectral classification, is not sensitive to gravity and is 
therefore similar in dwarfs, giants and young stars with the same spectral 
type; averaging the dwarf and giant spectra does not significantly change this 
particular feature.  The dwarf and giant standards averaged and used to 
compare with the OMC 2/3 brown dwarf candidates are referred to throughout 
this paper as luminosity class IV \citep*[and are from][]{l97,l98,l99}.

As was done by \citet{l99}, the OMC 2/3 spectra were compared with the 
template dwarf spectra, giant spectra, and the average of a dwarf and a giant 
spectra.  Reddening was also applied to the standards to duplicate the shape 
of the observed OMC 2/3 sources.  The spectral type and luminosity class that 
provided the best match as judged by visual inspection was adopted.  Since 
molecular features change rapidly with temperature, the resulting 
uncertainties in spectral type are $\pm$ 0.25 subclass. 

Tables \ref{table:browndwarfs}, \ref{table:lowmassmemberspectra} and 
\ref{table:fieldspectra} summarize the results of the spectroscopic 
observations.  The tables include: those objects determined to be young brown 
dwarfs with spectral types later than M6.5 (Table \ref{table:browndwarfs}), 
young, low mass members with spectral types of M4$-$M6 (Table 
\ref{table:lowmassmemberspectra}) and field dwarfs or giants and sources whose 
spectral types could not determined (Table \ref{table:fieldspectra}).  

\begin{figure} 
\epsscale{.80}
\plotone{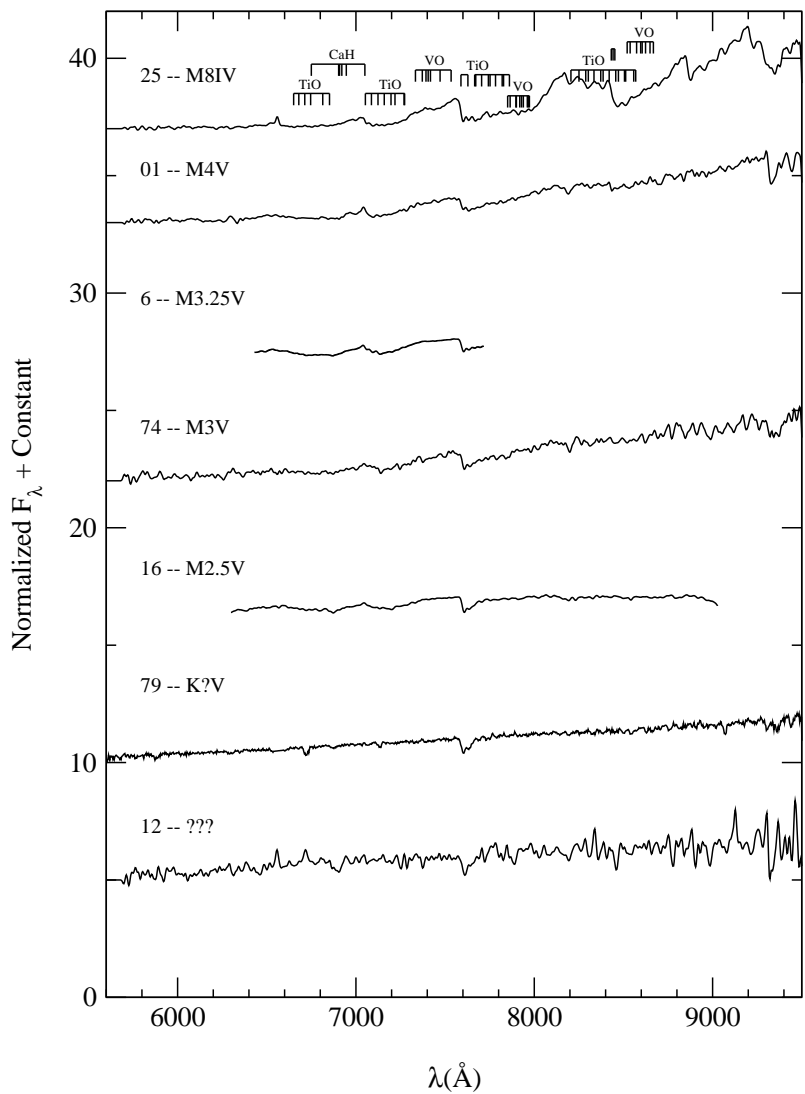}
\caption[Far-Red Spectra]{Low resolution spectra obtained with LRIS 
(6400$-$7700 \AA) on Keck I as well as with the Red (5500$-$9800 \AA) and Blue 
(6200$-$9000 \AA) Channel Spectrographs on the MMT of 
candidate brown dwarfs in OMC 2/3 that are either field dwarfs, or whose 
spectral types are undetermined (from Table \ref{table:fieldspectra}).   Note 
that candidate 25 is indeed a young brown dwarf, and has been classified as 
such; however, this source was serendipitously observed and since we are not 
confident we have the correct source due to uncertainties in reconstructing 
its position, it is grouped with the unclassified sources.  The spectra have 
been smoothed to a resolution of 18 \AA~ and normalized at 7500 \AA.  }
\label{fig:mmtspectra3}
\end{figure}

Of the seven candidates observed with LRIS, three candidates are young M6.5 
members (see Figure \ref{fig:staufspectra}), three were determined 
to be low mass stars in OMC 2/3, and one is a field dwarf, best matching a 
M3.25 dwarf spectrum (see Table \ref{table:fieldspectra}).  For the MMT Blue 
Channel Spectrograph observations, of the twelve candidates observed, three 
were determined to be young with spectral types ranging from M7$-$M9; 
therefore, they are young, brown dwarfs.  Eight of the remaining nine are low 
mass stars with spectral types ranging from M5$-$M6 and the ninth was 
determined to be a M2.5 field dwarf.  Finally, of the thirteen candidates 
observed with the MMT Red Channel Spectrograph, three of them are young brown 
dwarfs with spectral types later than M6.5 (they are also shown in Figure 
\ref{fig:mmtspectra}), five are young, low mass stars with spectral types 
ranging from M4$-$M6, and the remaining five are undetermined.  Figure 
\ref{fig:mmtspectra2} shows the far-red spectra of the remaining M4$-$M6 
members and Figure \ref{fig:mmtspectra3} shows the remaining sources that 
appear in Table \ref{table:fieldspectra}.

\placefigure{fig:staufspectra}

\placefigure{fig:mmtspectra}

\placefigure{fig:mmtspectra2}

\placefigure{fig:mmtspectra3} 

\subsubsection{H$\alpha$ Emission \label{sec:halpha}}

Strong H$\alpha$ emission is a distinguishing feature of many young pre-main 
sequence stars, and similarly young substellar objects 
\citep[e.g.][]{muzerolle05}, and its presence in a stellar spectrum is an 
indication of accretion onto a star from the surrounding disk.  In general, 
the strength of H$\alpha$ emission defines the difference between weak-line T 
Tauri stars and classical T Tauri stars, i.e. it indicates whether an object 
is actively accreting material or not.  Line strength is measured here by its 
equivalent width which indicates whether the H$\alpha$ emission is produced 
from gas accretion or originates from chromospheric activity.  Traditionally, 
a WTTS is defined as having EW[H$\alpha$] $>$ $-$10 \AA~ and an EW[H$\alpha$] 
$<$  $-$10 \AA~ is considered to be an actively accreting classical T Tauri 
star (CTTS).  However, here we will use an updated definition of a CTTS, 
developed by \citet{rw03} to have EW[H$\alpha$] $\leq$ $-$20 \AA~ for those 
objects with M3$-$M5.5 spectral types and EW[H$\alpha$] $\leq$ $-$40 \AA~ for 
those with M6$-$M7.5 spectral types.  Although stars with EW[H$\alpha$] $>$ 
$-$40 \AA~ may still be accreting, we cannot distinguish between accretion and 
chromospheric activity without resolving the H$\alpha$ line and measuring its 
velocity width.

The best examples of H$\alpha$ emission at 6563\AA~ can be seen in all three 
far-red spectra in Figure \ref{fig:staufspectra} and in the spectra of 
candidates 17 and 21 in Figure \ref{fig:mmtspectra}.  In particular, candidate 
21 exhibits an equivalent width for H$\alpha$ stronger than $-$100 \AA, 
which with a M7 spectral type indicates that it is an actively accreting young 
brown dwarf.  Based on the \citet{rw03} definition, 
however, the three M6.5 brown dwarfs that show H$\alpha$ emission in Figure 
\ref{fig:staufspectra} have EW[H$\alpha$] greater than $-$40 \AA~ and may not 
be accreting.  Table \ref{table:browndwarfs} indicates the H$\alpha$ 
equivalent widths for all the spectroscopically confirmed brown dwarfs.  

There are five low mass members with spectral types ranging from M4$-$M6 which 
also show detectable H$\alpha$ emission (for values of EW[H$\alpha$], see 
Table \ref{table:lowmassmemberspectra}).  Three of the five have 
EW[H$\alpha$] $\leq$ $-$20 \AA, therefore they can be considered to be 
classical T Tauri stars.  The remaining two are likely weak-line T Tauri stars 
based on their equivalent widths, but to be sure, the \ion{Li}{1} (at 6708 
\AA) absorption abundance should be known \citep[e.g.][]{martin97}.  Higher 
resolution spectra would be useful for confirmation of this distinction.

\citet{rw03} note that low resolution spectra such as the spectra discussed 
here often overestimate the equivalent width of H$\alpha$ for mid-M and cooler 
objects.  Because the feature is located near the 6569 \AA~TiO absorption 
band, at low resolution the H$\alpha$ feature can combine with the edge of the 
TiO band on one side leading to the underestimation of the continuum, and 
therefore an overestimation of the equivalent width.  In the case of candidate 
21, EW[H$\alpha$] $< -$40 \AA, and its identification as an actively 
accreting young brown dwarf appears to be secure.

\subsubsection{\ion{Li}{1}, \ion{K}{1}, \ion{Na}{1}, \& \ion{Ca}{2} Absorption \label{sec:lithium}}

Several atomic absorption features in the optical can be used to 
establish the youth of the observed candidates: these include \ion{Li}{1}, 
\ion{K}{1}, \ion{Na}{1}, and \ion{Ca}{2}.  The detection of lithium provides 
verification of youth for pre-main sequence stars or verification of 
substellar mass for field dwarfs.  In most of our far-red spectra, lithium is 
not detected due to the low spectral resolution and modest signal-to-noise of 
our spectra.  The Keck LRIS spectra show \ion{Li}{1} in two of the stellar 
sources (candidates 5 and 6.n in Table \ref{table:lowmassmemberspectra}) 
confirming the youth of these two pre-main sequence stars.  However, for the 
faint young brown dwarfs observed with Keck (candidates 1, 2 and 3 in Table 
\ref{table:browndwarfs}), the signal-to-noise ratio was too low for 
\ion{Li}{1} detection.  In addition, \citet{kirk06} caution that the strength 
of the \ion{Li}{1} line is gravity dependent and therefore is not a useful 
test of the substellar nature of young brown dwarfs.

In the spectra obtained with the Red and Blue Channel Spectrographs at the 
MMT,  \ion{K}{1} ($\sim$7700 \AA) and \ion{Na}{1} ($\sim$8200 \AA) absorption 
are detected and these lines are good for discriminating young members 
of OMC 2/3 from field dwarfs.  Both lines are sensitive to gravity, 
both being much weaker in young, low gravity objects than in higher gravity 
field dwarfs \citep[as compared in Figure 2 of][]{l03}.  The spectra in Figure 
\ref{fig:mmtspectra} show that both absorption features are extremely weak 
when detected in the probable OMC 2/3 objects identified in Section 
\ref{sec:molecular}, further confirming that these are young stars.  In order 
to distinguish from background giants which have even lower surface gravities 
than young members, the \ion{Ca}{2} triplet ($\sim$8500$-$8660 \AA) can be 
used as a discriminator: strong \ion{Ca}{2} absorption is indicative of 
giants.  However, the \ion{Ca}{2} triplet is not detected in any of the 
spectra.

\subsection{Near-Infrared Spectra}

\subsubsection{SpeX Spectra}

There have been 56 brown dwarf candidates observed to date with SpeX; four of 
those sources were also observed in the far-red and discussed in the previous 
section.  The 10 members that have spectral types of M7$-$M9 are summarized in 
Table \ref{table:browndwarfs} and the 4 members with spectral types of M4$-$M6 
are summarized in Table \ref{table:lowmassmemberspectra}.  The remaining 
foreground/background non-members are summarized in Table 
\ref{table:fieldspectra} along with sources of indeterminate type for which 
spectra have been obtained.  Many of the objects listed in Table 
\ref{table:fieldspectra} have relatively featureless spectra (no detectable 
water absorption features) at this resolution ($\lambda$/$\Delta\lambda \sim$ 
250) and are likely reddened field dwarfs or giants.  In addition to the brown 
dwarf candidates, two dozen standards classified from far-red spectra were 
observed with SpeX for comparison with the candidates in order to quantify the 
shape of the water absorption features indicative of late M-type stars. 

\begin{figure}
\epsscale{1.0}
\plotone{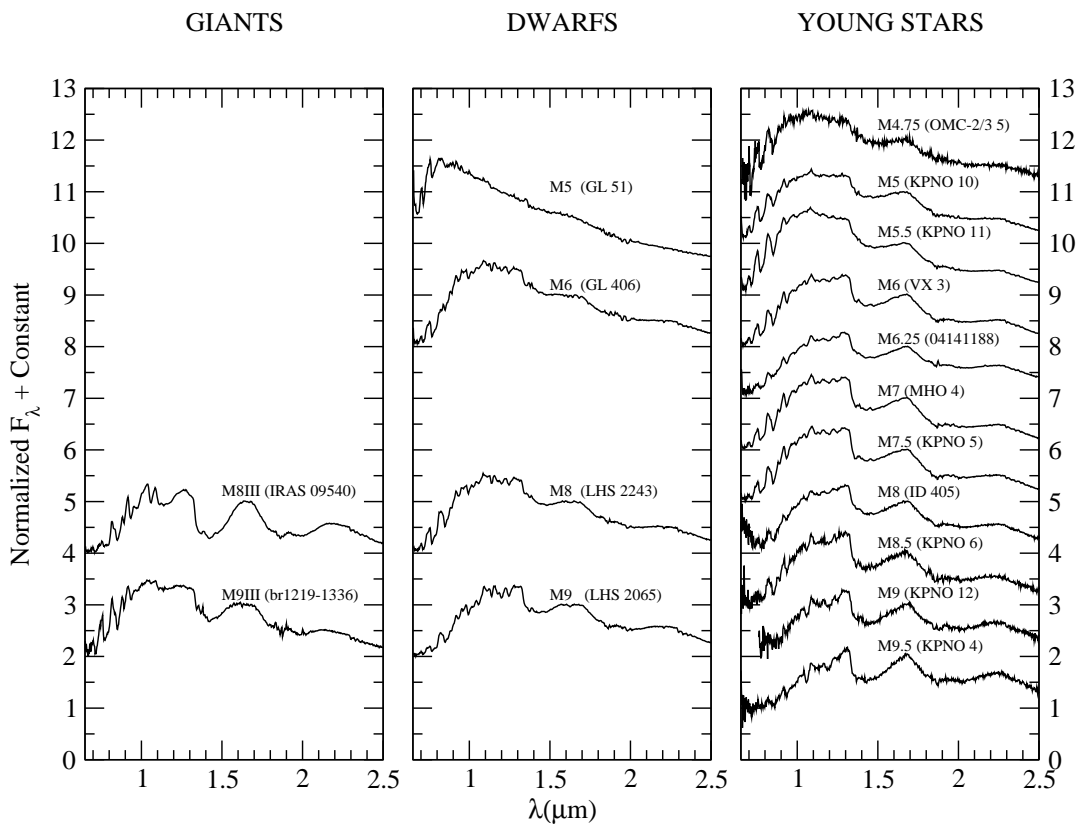}
\caption[SpeX Near-Infrared Spectra: Spectral Classification]{For 
comparison with the OMC 2/3 brown dwarf candidate near-infrared spectra, a 
grid of standards was also obtained with the SpeX Spectrograph in single-prism 
mode ($\lambda$/$\Delta\lambda \sim$ 250).  These objects were classified from 
their far-red spectra in \citet{briceno02} and \citet{l03}.  A representative 
sample of the SpeX spectra of these standards is presented here.  They were 
observed during the December 2003 run described in Section 
\ref{sec:spexspectra} as well as by \citet{luhman06} and \citet{m06ip}.  These 
spectra have all been normalized at a wavelength of 1.66 $\mu$m.}
\label{fig:spectypes}
\end{figure}

Near-infrared spectra of late M (and L) dwarfs are dominated by 
H$_2$O absorption bands at 1.4 and 1.85 \micron.  These steam absorption 
bands characterize late M dwarfs by broad, plateau-like features in the 
spectrum \citep{reid01,leggett01}.  However, young stars and brown dwarfs look 
quite different in that they exhibit sharp, triangular shaped peaks in the 
$H-$ and $K-$bands \citep{lrah01}.   The sharp peaks in the $H-$band are 
possibly due to a reduction in H$_2$ collision induced absorption 
\citep{kirk06}.  As shown in Figure \ref{fig:spectypes}, the shape of the 
$H-$band peak is strongly dependent on surface gravity, and appears to show 
distinct shapes for field dwarfs, pre-main sequence stars and giants.

\placefigure{fig:spectypes}

Figure \ref{fig:spectypes} displays a representative sample of the grid of 
pre-main sequence star and young brown dwarf standards used to classify the 
candidate brown dwarfs in OMC 2/3.  The grid of standards was compiled from 
Taurus and IC 348 spectra obtained with SpeX as part of this study (see 
Section \ref{sec:spexspectra}) as well as those obtained by \citet{luhman06} 
and \citet{m06ip}.  Each candidate spectrum was compared to each spectrum in 
the grid of standards, reddened using the following interstellar extinction 
law from \citet{draine89}:

\begin{equation}
A_\lambda \propto \lambda^{-1.75}.
\end{equation}

\begin{figure}
\epsscale{.80}
\plotone{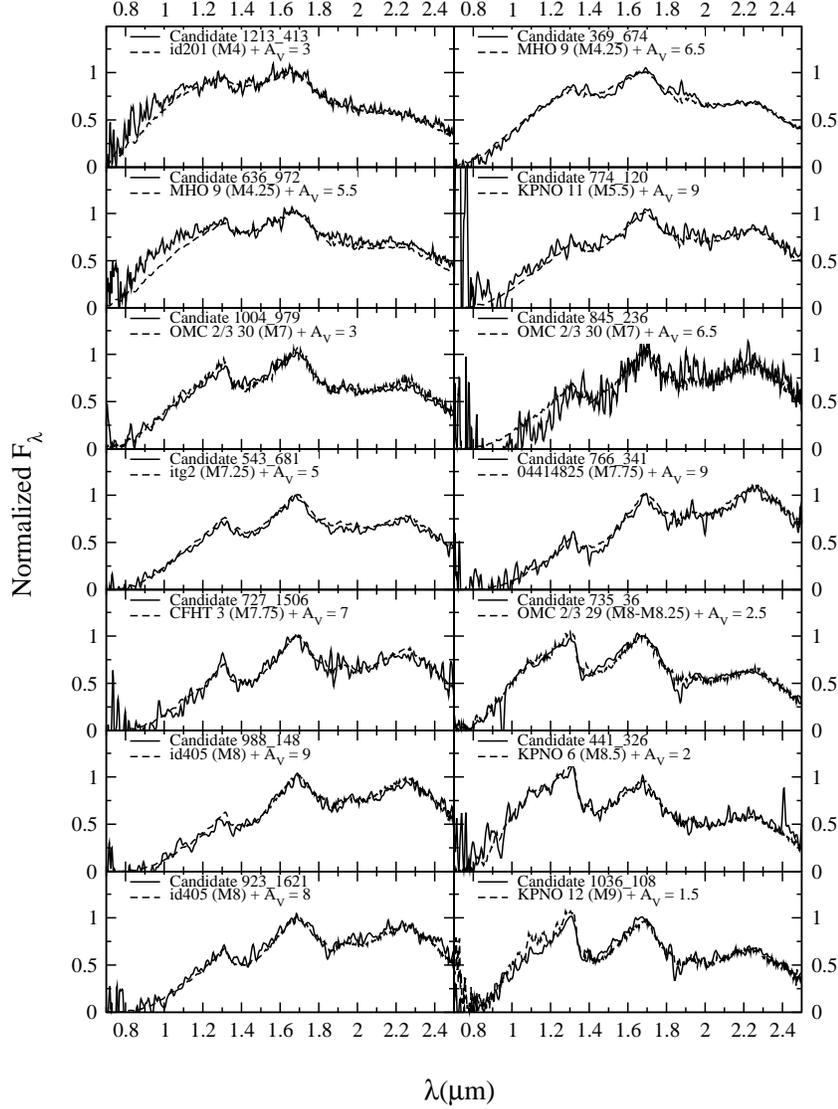}
\caption[SpeX Near-Infrared Spectra]{These 14 spectra of embedded OMC 2/3 
members (solid lines) were obtained at the IRTF with SpeX in single-prism mode 
($\lambda$/$\Delta\lambda \sim$ 250); for display purposes they have been 
smoothed with a Gaussian kernel.  The reddened reference spectra (dashed 
lines) with the best match to the candidates are over-plotted.  The reference 
spectra were also observed with SpeX, but their spectral types were obtained 
from far-red spectra.  Note the locations of the H$_2$O and CO absorption 
bands, the shapes of which were used for spectral classification.  All spectra 
were normalized at a wavelength of 1.68 $\mu$m.}
\label{fig:irspectra}
\end{figure}

A custom IDL routine computed the value of chi-square for each template 
reddened by a range of extinction values and selected the one that minimized 
chi-square as the best fit.  The \citet{draine89} extinction law was used, and 
although there is some variation in the value of the exponent of the 
extinction law in the literature, we find that the derived spectral types do 
not depend on the precise value of the exponent.  In addition, a particular 
spectral range can be chosen for the fit; however, in general most of the 
entire spectral range for SpeX was used for classification of these spectra: 
0.9$-$2.4 $\mu$m.  The result, shown in Figure \ref{fig:irspectra}, is a best 
fit standard spectrum (plus $A_V$), overplotted on the verified M4$-$M9 low 
mass members in OMC 2/3.

\placefigure{fig:irspectra}

\begin{figure}
\plotone{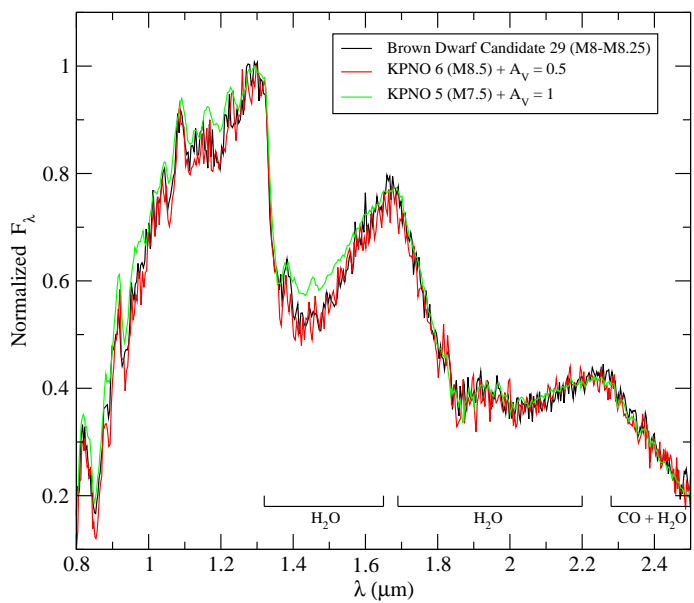}
\caption[SpeX Spectrum of Brown Dwarf Candidate 29]{The spectrum of brown 
dwarf candidate 29 (solid line) obtained at the IRTF with the SpeX 
Spectrograph in single-prism mode ($\lambda$/$\Delta\lambda \sim$ 250).  This 
brown dwarf was classified from its far-red spectrum (see Figure 
\ref{fig:mmtspectra}) as M8$-$M8.25 with A$_V$ = 0.  The dashed line (colored 
red in electronic edition) shows the SpeX reference spectrum from a young 
member in Taurus, KPNO 6, with a spectral type of M8.5 and given a reddening 
of A$_V$=0.5, which yields a spectrum very similar to OMC 2/3 brown dwarf 
candidate 29.  The spectrum from KPNO 5 (dotted line colored green in 
electronic edition), a Taurus low mass member with a known spectral type of 
M7.5 \citep{briceno02} is also shown; note that the OMC 2/3 brown dwarf 
candidate spectrum is a better match to the M8.5 spectrum than the M7.5 
spectrum (specifically when looking at the depth of the steam band from 
1.3$-$1.5 $\mu$m).  An uncertainty of $\pm$ 0.5 subclass is typical for these 
spectra.  All three spectra were normalized at a wavelength of 1.30 $\mu$m.  
See the electronic edition of the Journal for a color version of this figure.}
\label{fig:bd29illus}
\end{figure}

Figure \ref{fig:bd29illus} illustrates the triangular shaped peaks in the 
near-infrared spectrum of candidate 29 in OMC 2/3, which was classified as 
M8$-$M8.25 from its far-red spectrum.  The best fit spectral type from the 
near-infrared spectrum is M8.5 with an A$_V$ = 0.5.  Because of these 
triangular shaped features, it is relatively easy to distinguish between young 
members and foreground dwarfs or background giants that may contaminate the 
candidate sample.  By varying the spectral type of the standard being used for 
the fit (as shown in Figure \ref{fig:bd29illus}), we estimate an uncertainty 
in spectral class of $\pm$ 0.5 subclass.

\placefigure{fig:bd29illus}

\begin{figure}
\epsscale{.80}
\plotone{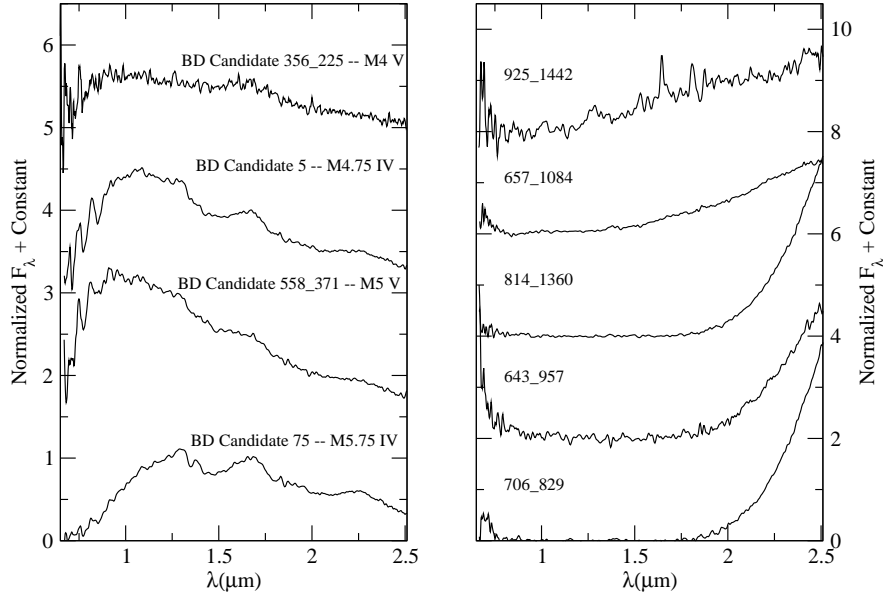}
\caption[SpeX Near-Infrared Spectra]{These 9 spectra were obtained at the IRTF 
with SpeX in single-prism mode ($\lambda$/$\Delta\lambda \sim$ 250); for 
display purposes they have been smoothed with a Gaussian kernel.  In the 
left-hand panel, we show the spectra that have been classified as either 
members (IV spectral class) or dwarfs (V spectral class).  Brown dwarf 
candidates 5 and 75 are young stars which have already been displayed in the 
far-red in Figures \ref{fig:mmtspectra2} and \ref{fig:mmtspectra}, 
respectively.  The sources in the right-hand panel all had indeterminate 
spectral types (Table \ref{table:fieldspectra}).  The spectra in the left-hand 
panel were normalized at a wavelength of 1.68 $\mu$m, and in the right-hand 
panel, they were normalized at a wavelength of 2.2 $\mu$m.  The photometry for 
these sources appears in Tables \ref{table:lowmassmemberspectra} and 
\ref{table:fieldspectra}.}
\label{fig:irspectra2}
\end{figure}

\begin{figure}  
\epsscale{.80}
\plotone{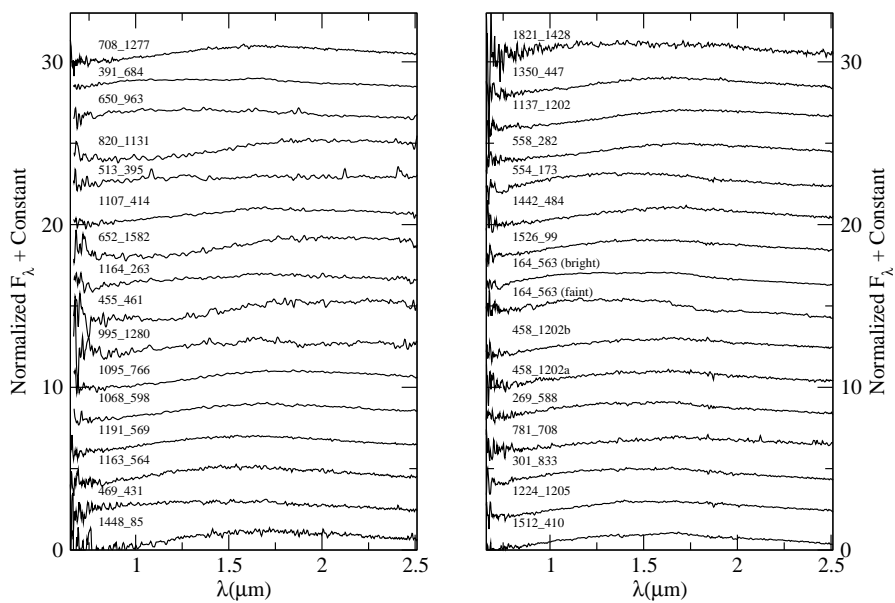}
\caption[SpeX Near-Infrared Spectra]{These 32 spectra, all sources with 
indeterminate spectral types (Table \ref{table:fieldspectra}), were obtained 
at the IRTF with SpeX in single-prism mode ($\lambda$/$\Delta\lambda 
\sim$ 250); for display purposes they have been smoothed with a Gaussian 
kernel.  The photometry for these sources appears in Table 
\ref{table:fieldspectra}.  All spectra were normalized at a wavelength of 1.68 
$\mu$m.}
\label{fig:irspectra3}
\end{figure}

In addition to candidate 29, several other low mass members, specifically 
candidates 30, 5 and 75, were observed in the far-red and with SpeX.  For 
candidate 30, the best fit near-infrared type is M7.25 and the far-red type, 
M7.  Candidate 5 has a near-infrared type of M4.75 and a far-red type of 
M4.75; both spectra indicate that this is a low mass member of OMC 2/3.  
Candidate 75 has a best fit near-infrared type of M5.75$-$M6 and the far-red 
type is M6.  In general, near-infrared spectral types of candidates with 
existing far-red spectra give spectral types within 0.5 subclass of the 
far-red spectral type.  The spectra for candidate 30 is shown (as a spectral 
template for sources 1004\_979 and 845\_236) in Figure \ref{fig:irspectra} and 
candidates 5 and 75 are shown in Figure \ref{fig:irspectra2}.  Spectra for all 
the sources of indeterminate type (see Table \ref{table:fieldspectra}) 
observed with SpeX can be found in both Figures \ref{fig:irspectra2} and 
\ref{fig:irspectra3}.

\placefigure{fig:irspectra2}
\placefigure{fig:irspectra3} 

\subsubsection{CorMASS Spectra}

\begin{figure}
\plotone{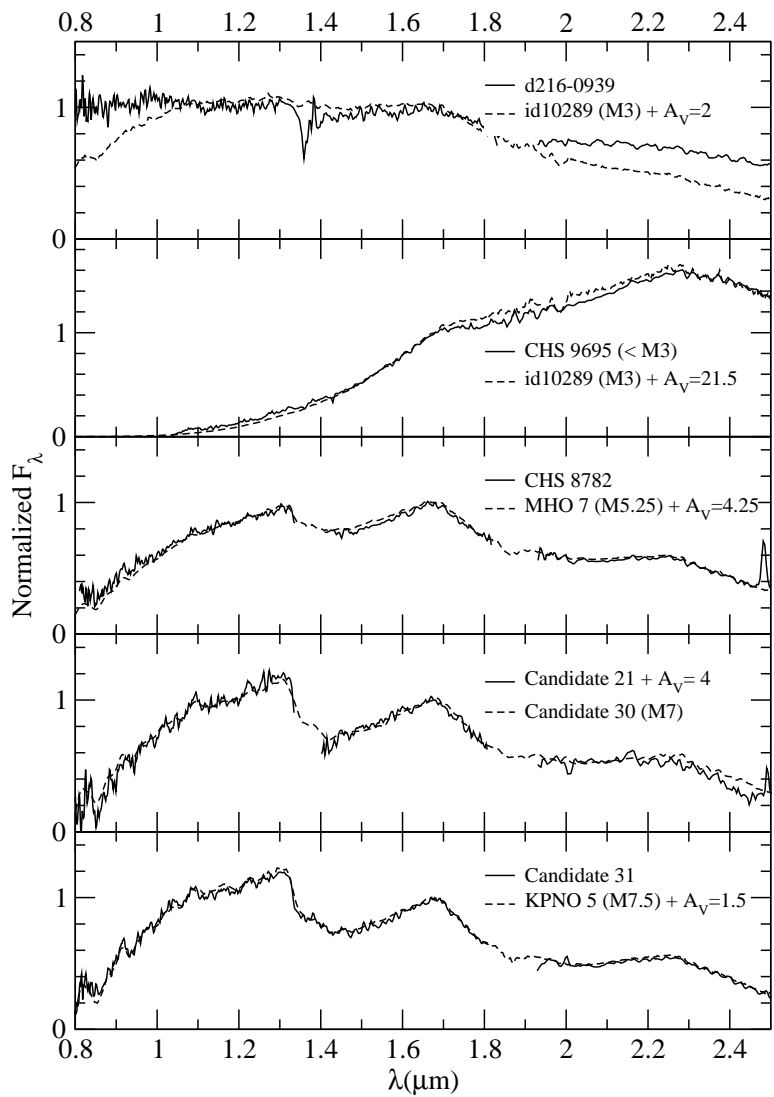}
\caption[CorMASS Near-Infrared Spectra]{These 5 spectra (solid lines) 
were obtained with CorMASS ($\lambda$/$\Delta\lambda \sim$ 300); for display 
purposes they have been smoothed with a Gaussian kernel and some regions where 
telluric was very noisy have been removed.  Reference spectra obtained with 
SpeX (dashed lines) with the best match to the candidates are over-plotted 
(spectral types determined from far-red spectra).  Note that candidate 21 
(obtained with CorMASS) is virtually identical to candidate 30 (obtained with 
SpeX); both candidates were classified as M7 from far-red spectra.  All 
spectra have been normalized at 1.68 $\mu$m.}
\label{fig:cormassspectra}
\end{figure}

Spectral classification for the CorMASS spectra used the same method as 
for the SpeX spectra, though the regions of the spectra dominated by 
absorption from telluric water vapor were noisier in the CorMASS spectra and 
so were removed during the classification process.  Figure 
\ref{fig:cormassspectra} shows five of the six near-infrared spectra obtained 
with CorMASS; in two of the cases the spectra obtained were for sources 
already classified from far-red spectra.  Candidate 31, which we found to have 
a far-red spectral type of M7.75, has a near-infrared spectral type from 
CorMASS data of M7.5.  For the other source, candidate 21, the CorMASS and 
far-red spectral types are also within 0.25 subclasses.

To demonstrate that the CorMASS spectra can be classified using the SpeX 
standard spectra, we compared candidates 21 and 30, both of which have far-red 
spectral types of M7.  Candidate 21 was observed with CorMASS and candidate 30 
was observed with SpeX. As illustrated in Figure \ref{fig:cormassspectra}, 
these spectra have virtually identical shapes: candidate 30 looks like 
candidate 21, reddened with an A$_V$=4.  Therefore we maintain that the SpeX 
spectra and CorMASS spectra are comparable, within 0.25 subclasses.  It is 
useful to note that this was also seen in \citet{letal06} and \citet{allers07}.

\placefigure{fig:cormassspectra}

The remaining spectra obtained with CorMASS that did not overlap with previous 
far-red or SpeX observations are also illustrated in Figure 
\ref{fig:cormassspectra}.  The source d216-0939 was imaged by \citet{sblw05} 
with Advanced Camera for Surveys on the Hubble Space Telescope as a nearly 
edge-on disk.  Its near-infrared spectrum, shown in Figure 
\ref{fig:cormassspectra}, is shown with a young M3 star over-plotted.  It is 
clear that the spectral type of this source is earlier than M3.  However, 
since we do not have template near-infrared spectra for young stars with 
spectral types earlier than M3, we were unable to precisely determine the 
type.  In addition, when we compared d216-0939 with field dwarfs earlier than 
M3, it was difficult to match the shape of these sources with the shape of 
d216-0939 because of a near-infrared excess which we attribute to the disk.  
Therefore, this object is classified as having an indeterminate type and is 
listed in Table \ref{table:fieldspectra}.

The CorMASS spectrum of CHS 9695 indicates that it is a highly reddened object 
(see Figure \ref{fig:cormassspectra}).  Although the spectrum very closely 
matches a spectrum of a young M3 object, it is slightly shallower and 
therefore, as with d216-0939, we were also unable to precisely determine its 
type (see Table \ref{table:fieldspectra}).  The candidate CHS 8782 was 
successfully classified from its CorMASS spectrum through our comparison 
method as a member of OMC 2/3 with a spectral type of M5.25.  One additional 
candidate, 583\_352 has a CorMASS spectrum with very low signal-to-noise and 
therefore we are unable to confidently determine a spectral type.

\section{Discussion \label{sec:omc23age}}

Spectra have been obtained for 88 candidate low mass members in OMC 2/3, of 
which 40 new low mass members with spectral types of M4$-$M9 have been 
identified.  Of those, 19 have spectral types later than M6.5, making them 
{\it bona fide} young brown dwarfs.  As seen in Figure \ref{fig:image_dist}, 
the spatial distribution of the young brown dwarfs as compared with the low 
mass M4$-$M6 sources is fairly similar.  

We have observed all the brown dwarf candidates that fall below 0.080 
$M_{\odot}$ and above 0.050 $M_{\odot}$ (for 1~Myr objects), and with an 
extinction limit of $4.5 < A_V < 14$, making a complete sample within this 
regime (see Figure \ref{fig:irtfpropfigs2003}).  Within these parameters lie 
a total of 22 candidates and of those, 11 are spectroscopically confirmed 
members with spectral types of M4$-$M9, and 11 are objects of indeterminate 
type (one of which, 583\_352, might be a member but a higher signal-to-noise 
spectrum is necessary for confirmation).  Therefore we find that $\sim$50\% of 
candidates detected within this range of extinction and dereddened magnitude 
are members.

\subsection{The H-R Diagram \label{sec:lbol}}

In order to determine the masses and ages of our sample, we estimate the 
effective temperatures and bolometric luminosities for the 40 low mass members 
and place them on an H-R diagram.  The effective temperature, T$_{eff}$, for 
each source was determined from its spectral type using the temperature scale 
of \citet{l03}.  We used the \jh~color to determine the reddening and 
dereddened the $H-$band magnitude; then the bolometric correction was 
applied.  The methodology is described in detail in Appendix 
\ref{sec:lbol_teff}.  

It is important to note that the A$_V$ values used to compute the dereddened 
$H-$band magnitude are not based on the ones found from spectral 
classification, but are based on the \jh~color.  The reason for this is that 
the standard spectra used for comparison are the {\it observed spectra} and 
have not been dereddened.  Although the A$_V$ values for these sources are 
low (likely A$_V <$ 2), we wanted to deredden all the spectra uniformly and 
therefore chose to do so using the \jh~color.

Figure \ref{fig:hrdiagram} shows the resulting effective temperatures and 
bolometric luminosities for the OMC 2/3 members on an H-R diagram (note that 
only 39 of the 40 low mass members appear on the diagram because source 
774\_120 does not have a $J-$band magnitude and so its bolometric luminosity 
could not be calculated in the same way as the others).  Over-plotted are the 
1, 3, and 10~Myr isochrones from \citet{bcah98}.  These isochrones are 
appropriate since we have adopted the temperature scale of \citet{l03} which 
has been calibrated for use with these isochrones.  Based on these 
evolutionary tracks, we estimate the 40 members with M4$-$M9 spectral types to 
range in mass from 0.35$-$0.015 $M_{\odot}$.  In general, though, the 
evolutionary models at present do not extrapolate consistent masses for 
sources with masses below the hydrogen burning limit \citep{hw04}.   An 
additional uncertainty in the H-R diagram of OMC 2/3 lies in the adopted 
distance of 450 pc.  Using the more recent value of 420 pc 
\citep[e.g.][]{mrfb07} instead of 450 pc shifts the points down on the H-R 
diagram, to later ages, by a value of $\log$ (L$_{bol}$/L$_\odot$) $\sim$ 0.06.

\placefigure{fig:hrdiagram}

Seventeen of the nineteen sources with spectral types later than 
M6.5 are located above the 1~Myr track.  However, two of them (candidates 21 
and 845\_236) sit well below the 3~Myr track.  For the 21 members with 
spectral types between M4 and M6, 11 of them fall either very close to the 
1~Myr isochrone or between the 1 and 3~Myr isochrones, and there are 8 likely 
members that fall very close to, or below the 3~Myr isochrone.  Although there 
are several sources that appear clustered at an age below the 3~Myr isochrone 
on the H-R diagram, this is primarily due to a selection effect from our 
requirement that the luminosity of our candidates be less than that of a 
1~Myr brown dwarf.  Without spectral types for the more luminous sources 
in OMC 2/3, we cannot determine whether these ages are typical, or 
whether most of the stars are indeed above 1~Myr in age.  We have detected 
one M4 source above the 1~Myr isochrone; this source serendipitously fell 
into the slit during our observations and was not in the sample selected 
from our photometry.
 
Also shown in Figure \ref{fig:hrdiagram} are the CorMASS sources CHS 9695, 
a source which is variable (and hence a likely member), as seen by 
\citet*{chs01}, and d216-0939, the edge-on disk discovered by \citet{sblw05}.  
Since we have assigned an upper limit of M3 for the spectral types for these 
two objects (see Table \ref{table:fieldspectra}), the locations of these 
sources on the H-R diagram are plotted for spectral classifications of M1, M2 
and M3.  CHS 9695 is clearly young, falling well above the 1~Myr track for all 
three spectral types.  The edge-on disk, d216-0939, falls between the 3~Myr 
and 10~Myr tracks for all three spectral types.  We caution that the 
possibility of strong veiling for d216-0939 in the 1$-$1.8 $\mu$m regime could 
result in a significant overestimation in the temperature of this object.

The fact that d216-0939, as well as candidates 21 and 845\_236 mentioned 
above, appear older than 3~Myr on the H-R diagram suggests that it is possible 
these sources are underluminous and young but appear older on the H-R diagram 
due to an obscuring disk.  Disks may absorb the light of the star, 
particularly if they are highly flared \citep{walker04}; however, the star may 
not appear unusually red due to a strong contribution of light scattered off 
the disk.  Hubble Space Telescope images of d216-0939 show a strong 
contribution from scattered light in this system \citep{sblw05}.  In support 
of this, candidate 21 shows strong H$\alpha$ emission, suggesting that it has 
an accretion disk.  

To obtain a more accurate estimate of the age of OMC 2/3, 
we need spectra of candidate members with types earlier than M6 and with 
luminosities above our luminosity limit to fill in that portion of the H-R 
diagram.  With the H-R diagram we currently have, we can make the 
following observations.  First, eighteen of the forty OMC 2/3 sources (45\%), 
and in particular, seventeen of the nineteen (89\%) young brown dwarfs, are 
located above the 1~Myr isochrone.  Therefore we estimate an age for OMC 2/3 
of approximately 1~Myr.  Second, although many of the members are close to or 
above the 1~Myr isochrone, we clearly detect a range of ages along the 
line-of-sight to OMC 2/3.  We now examine the implications of this observed 
age spread.

\placefigure{fig:avhisto}

\subsection{Age Spread \label{sec:agediscussion}}

The wide range of inferred ages detected in OMC 2/3 is surprising given the 
high density of protostars, and hence the presumed youth, of this region.  
There has already been considerable debate on the nature of age spreads in the 
Orion molecular cloud.  \citet{hs06} find evidence for an age spread of 
8$-$10~Myr in the H-R diagram of stars in the ONC, which they argue is real.  
\citet{ps99} find evidence of an increasing (i.e. accelerating) rate of star 
formation in the ONC which continued to rise until the present time.  In their 
age histogram for stars in the ONC (with masses ranging from 0.4$-$0.6 
$M_{\odot}$), all stars are less than 10~Myr in age, with most having ages 
less than 7~Myr.  \citet{h01}, and more recently \citet{hbw07}, argue that 
such age spreads may result from uncertainties in the luminosity, membership, 
and the pre-main sequence star tracks.  In addition, \citet{lodieu08} 
discuss how the presence of multiple systems can result in an age spread in  
their study of the Upper Sco association.  Multiplicity would result in an 
{\it underestimate} of the ages.  This likely effects only a very small 
percentage of sources in OMC 2/3 since the companion star fraction in the 
outskirts of the ONC for low mass PMS sources \citep[$\lae$5\%;][]{k06} has 
been found to be much lower than in other star-forming regions (a factor of 
5 lower than in Upper Sco).

In a study of low mass stars and brown dwarfs in the ONC, \citet{shc04} see 
two populations toward the ONC: a young, $\lae$ 1~Myr population and an older, 
10~Myr population.  However, in a more recent study of 45 new members of the 
Trapezium, an older, 10~Myr population of stars was not seen \citep{rrl07}; 
they suggest this may be the result of their source selection.  Although we do 
not see a clear bifurcation into two populations as in \citet{shc04}, our H-R 
diagram for the neighboring OMC 2/3 region shows a significant age spread as 
well.  Understanding the age spread in OMC 2/3 can bring new insight into the 
debate over the age spreads observed in Orion as a whole.

It is important to note that we rule out that the age spread seen is due to an 
error in spectral classification.  All candidates determined to be members in 
OMC 2/3 show low surface gravity features in their spectra that rule out the 
possibility that they are foreground M dwarfs.  In addition, the photometric 
uncertainty is small and any variability is likely on the order of tenths of 
magnitudes, not several magnitudes as would be required to change the H-R 
diagram significantly.  One possible source of uncertainty lies in our method 
of using a constant intrinsic \jh~color to calculate the color excess for each 
source (see Appendix \ref{sec:lbol_teff}); however, to test this, we plotted 
the same H-R diagram using the intrinsic \jh~values for each spectral type 
presented in \citet[][using the young disk population types found in their 
Table 6]{leggett92}.  Using the values presented in that table instead of a 
constant intrinsic \jh~color for all spectral types shifts the location of the 
sources with spectral types of M7.5 and later slightly closer to the 1~Myr 
isochrone, but does not shift them to ages older than 1~Myr.

We consider here several reasons for the spread of ages seen in OMC 2/3: 
confusion with members of the older subgroups in the Orion OB~1 association, 
uncertainties in the determination of luminosity and spectral type due to 
extinction from a disk, veiling, and finally, the presence of a {\it bona 
fide} age spread.

First, we assess the possibility of contamination from older members of the 
OB~1 association by examining the number of sources in each age range versus 
the amount of extinction toward that object (see Figure \ref{fig:avhisto}).
All 40 low mass members with spectral types ranging from M4$-$M9 are included, 
as well as the two CorMASS sources with upper limits to their spectral types 
of M3.  The sources are grouped into three categories: those with ages on the 
H-R diagram $<$ 1~Myr, those with ages 1$-$3~Myr, and those with ages $>$ 
3~Myr.  The low extinction values for the 1$-$3~Myr sources may indicate we 
are seeing a foreground population; or perhaps we are seeing the older, Orion 
OB 1c association which is 4.6~Myr old \citep{b94} and falls along the 
line-of-sight of OMC 2/3.  However, it seems unlikely that the sources greater 
than 3~Myr are foreground since they are, for the most part, more embedded.  
We cannot completely rule out this scenario because some of the older stars 
may show extinction due to circumstellar dust or from dust located within the 
Orion OB association but outside the molecular cloud.  Furthermore, some older 
stars could be trapped within the Orion molecular cloud, as suggested by 
\citet{pk07}.

Second, as discussed before, a combination of extinction and scattering from 
a nearly edge-on, flared circumstellar disk could result in the luminosity of 
a star being underestimated.  Since this would depend on the presence of a 
disk, as well as its inclination, the effect of disks could result in both 
increased scatter and an overestimated age.  The $>$ 3~Myr sources 
d216-0939, candidate 21, and candidate 845\_236 are possible examples of 
sources which appear underluminous due to the presence of a disk; these 
examples make this perhaps the most likely of these scenarios. As additional 
support for this scenario, at least three of the $\sim$10~Myr sources in the 
ONC are known proplyds or silhouette disks, and therefore younger than their 
positions on the H-R diagram suggest \citep{shc04,rrl07}.

Third, excess emission from nonphotospheric sources in the $J-$ and $H-$bands 
may effect our estimates of both luminosity and effective temperature.  
\citet{cieza05} find that for classical T Tauri stars, $J-$ and $H-$band 
luminosities are consistently overestimated, leading to an underestimation in 
age.  If we are to assume that there is significant $J-$ and $H-$band excess 
in our OMC 2/3 sources which we have not accounted for, then this would imply 
that we have overestimated the photospheric luminosity of the sources and that 
the ages are systematically underestimated.  This could increase the observed 
age spread by shifting sources with higher accretion rates to younger inferred 
ages; however, this would also imply that many sources are older than they 
appear on the H-R diagram.  In this case the mass, particularly of the brown 
dwarfs, would stay relatively constant since the mass tracks are virtually 
vertical at these ages (see Figure \ref{fig:hrdiagram}).  Near-infrared excess 
could also cause veiling and potentially bias sources to earlier types by 
reducing the observed depth of the water features.  Although we find some 
evidence for $K-$band excesses in the color-color diagram, we find no evidence 
for excesses in the spectra given the excellent fits in the near-infrared 
spectra and the consistency between the far-red and near-infrared spectral 
types.  Therefore, there is no clear evidence for veiling and significant 
excess emission in the $J-$ and $H-$bands.

Finally, it is possible that the age spread we see in OMC 2/3 is real.  This 
is the conclusion drawn by \citet{prfp05} based on their study of lithium 
depletion in the ONC.  They find low mass members of the ONC with ages $\gae$ 
10~Myr, which is much older than many of the stars in this region, and 
indicate that this may mean the star formation history of the cluster has a 
much longer duration than previously thought.  \citet{j07} used rotation 
periods and equatorial velocities of known PMS stars in the ONC to model the 
distribution of their projected radii and in turn, the distribution of ages.  
The simulations showed that an age spread of 1$-$3 Myr in the low-luminosity 
ONC members is consistent with the data, independent of the evolutionary 
models used.  However, the older stars discovered in the conventional H-R 
diagram analysis were not present in his analysis.   Jeffries concluded that 
he could not verify the presence of the 10~Myr sources.

In order to determine which of these possibilities is most likely in OMC 2/3, 
future studies should examine the ages of the stars above the hydrogen burning 
limit to see if most stars in OMC 2/3 have ages around 1~Myr, similar to the 
brown dwarfs, or if an age spread is seen in the higher mass PMS stars as 
well.  With such a complete sample, we could potentially fully understand 
the star formation history of OMC 2/3.

\section{Conclusions}

Forty new low mass members with spectral types of M4$-$M9 have been 
spectroscopically confirmed in OMC 2/3.  Using deep, \izjhk photometry of a 
20$^{\prime}$ $\times$ 20$^{\prime}$ field in OMC 2/3, we selected candidates 
with luminosities and colors consistent with brown dwarfs for follow-up 
spectroscopy.  Low resolution, far-red and near-infrared spectra were obtained 
for a sample of candidates and 19 objects with spectral types of M6.5$-$M9 and 
21 objects with spectral types of M4$-$M6 were confirmed by fitting these 
spectra to standard spectra, and presented.  These spectral types correspond 
to masses of 0.35$-$0.015 $M_{\odot}$ using the evolutionary models of 
\citet{bcah98} and we estimate an age for OMC 2/3 from the substellar members 
of approximately 1~Myr.  However, a significant fraction of selected low 
luminosity sources are pre-main sequence stars with ages ranging from 
1$-$10~Myr.

We discussed several reasons for the age spread we see in OMC 2/3.  First, 
that the older stars are part of the OB~1c association in the foreground to 
the Orion A cloud.  Although this possibility has not been ruled out, we find 
it unlikely because the older stars seem to be embedded in the Orion molecular 
cloud, as was found for the older stars in the ONC \citep{shc04}.  Second, 
that the measurement of the ages are effected by uncertainties in the 
luminosity of the sources.  Our data exhibit some evidence that the presence 
of disks can lead to overestimating the age of the sources.  For 
example, the apparent old ages of the edge-on disk d216-0939 and the strongly 
accreting candidate 21 may be explained by the presence of an edge-on or 
nearly edge-on flared disk which is obscuring the star.  Finally, it is 
possible that we are observing a {\it bona fide} age spread in OMC 2/3.

\acknowledgments

We are grateful to N. Caldwell for the Blue Channel Spectrograph observations 
and J. Rayner and B. Bus for help with SpeX observations.  We also thank M. 
Cushing for help with Spextool/Cormasstool and M. Merrill for providing some 
SQIID calibration data and help understanding SQIID linearity.  Finally, 
thanks go to the anonymous referee whose useful comments have improved this 
paper.  Support for STM was provided in part by NASA through contract 1256790  
issued by JPL/Caltech.  CorMASS is supported by a generous gift from the F.H. 
Levinson Fund of the Peninsula Community Foundation.  The authors also wish to 
recognize and acknowledge the very significant cultural role and reverence 
that the summit of Mauna Kea has always had within the indigenous Hawaiian 
community.  We are most fortunate to have the opportunity to conduct 
observations from this mountain.  This research has made use of NASA's 
Astrophysics Data System.

{\it Facilities:} \facility{ARC (CorMASS)}, \facility{FLWO:1.2m (4-Shooter)}, \facility{IRTF (SpeX)}, \facility{Keck:I (LRIS)}, \facility{KPNO:2.1m (SQIID)}, \facility{MMT (Red \& Blue Channel Spectrographs)}, \facility{VATT (CorMASS)}, \facility{Magellan:Clay (CorMASS)}

\appendix

\section{Bolometric Luminosity and Effective Temperature \label{sec:lbol_teff}}

Spectral types of stars and brown dwarfs correspond to an estimate of the 
atmospheric effective temperature (T$_{eff}$).  The specific effective 
temperatures are associated with spectral types (which we identify) and such a 
correspondence has been provided in the literature \citep{l03}.  Photometric 
observations are used to ascertain the bolometric luminosity of each star and 
brown dwarf candidate.  T$_{eff}$ and $L_{bol}$ are then compared against 
theoretical evolutionary models for a range of masses.

\begin{figure}
\epsscale{1.0}
\plotone{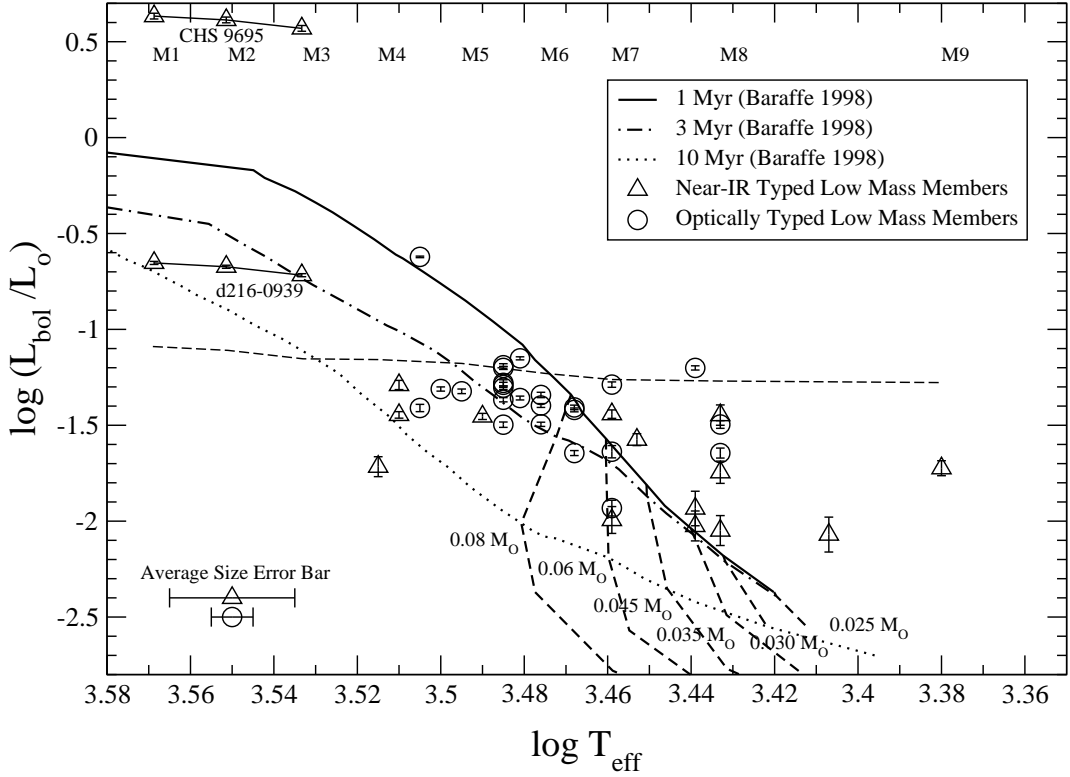}
\caption[OMC 2/3 H-R diagram: \citet{bcah98} Tracks]{A $\log$ 
(L$_{bol}$/L$_\odot$) versus $\log$ T$_{eff}$ H-R diagram for the 
spectroscopically confirmed low mass members of OMC 2/3.  Open circles 
indicate those whose spectral types were determined from far-red spectra and 
open triangles indicate those determined using near-infrared spectra.  
\citet{bcah98} evolutionary isochrones are shown for a 1 (top solid black 
line), 3 (middle dot-dashed line), and 10 (bottom dotted line) Myr-old PMS 
star.  The dashed lines indicate lines of constant mass, from left to right, 
0.08 M$_{\odot}$, 0.06 M$_{\odot}$, 0.045 M$_{\odot}$, 0.035 M$_{\odot}$, 
0.030 M$_{\odot}$ and 0.025 M$_{\odot}$.  The upper luminosity limit used to 
select the brown dwarf candidates is over-plotted (dashed line) to show the 
selection effect.  Two candidates observed with CorMASS, CHS 9695 and 
d216-0939 (the edge-on disk) are added here with their upper spectral 
classification limit of M3 to show where they appear on the diagram as an M1, 
M2 or M3 source.  Average error bars for uncertainty in $\log$ T$_{eff}$ are 
shown in the lower left-hand corner ($\pm$ 0.5 subclass uncertainty in 
spectral type for the NIR classified sources, and $\pm$ 0.25 subclass 
uncertainty for the optically classified sources).  The error bars shown for 
the $\log$ (L$_{bol}$/L$_\odot$) axis are based on photometric uncertainties 
from the $J-H$ colors.} \label{fig:hrdiagram}
\end{figure}

\begin{figure}
\plotone{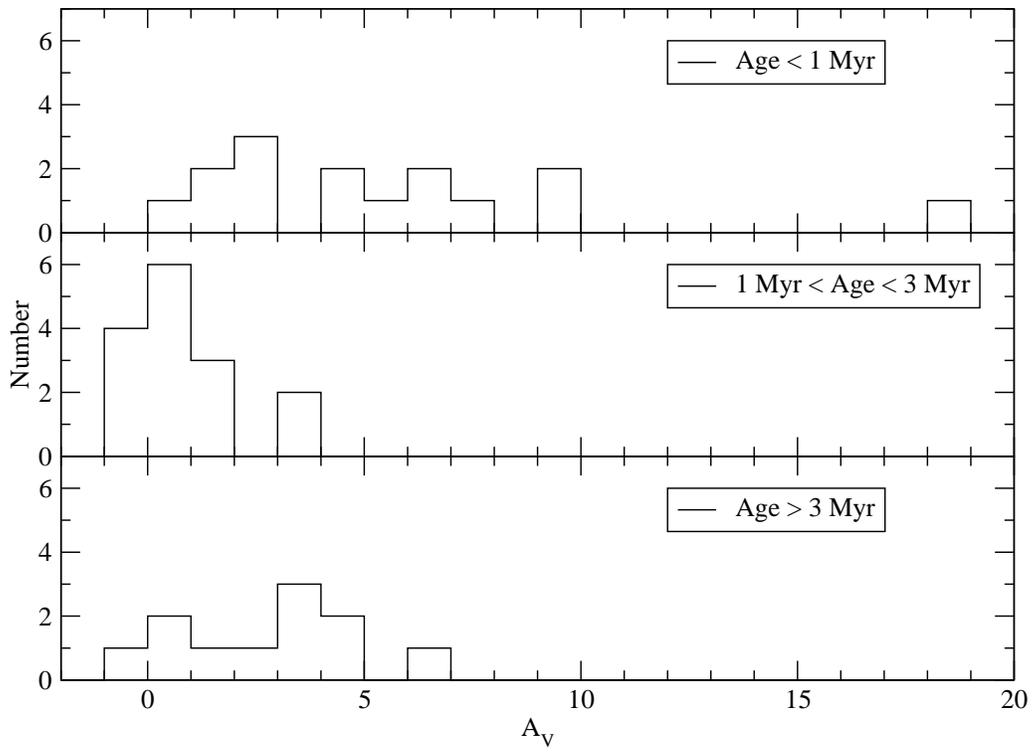}
\caption[A$_V$ Histogram]{Comparison of the values of A$_V$ 
for sources in three different age groupings: less than 1 Myr, between 1 and 
3 Myr and greater than 3 Myr.}
\label{fig:avhisto}
\end{figure}

In order to assess the bolometric luminosity for each source, we take the 
following steps: 

\begin{itemize}
\item Compute the excess in the $J-H$ color, $E(J-H)$, for each of the 
stars and brown dwarfs.

\begin{equation}
E(J-H) = (J-H) - (J-H)_{0}
\end{equation}

\noindent
where $(J-H)_{0}$ = 0.6.  This value is representative of the intrinsic $J-H$ 
colors given for young, M stars (ranging from M0$-$M9) in Table 6 of 
\citet{leggett92}.

\item The extinction, A$_H$ can be obtained from the color excess:

\begin{equation}
\frac{E(J-H)}{A_V} = \frac{A_J}{A_V} - \frac{A_H}{A_V} = 0.11
\end{equation}

\noindent
where values for A$_\lambda$/A$_V$ are from \citet{cfpe81}.  
Specific A$_H$ values for each of the candidates are listed in column 10 in 
Tables \ref{table:browndwarfs}, \ref{table:lowmassmemberspectra} and 
\ref{table:fieldspectra}.  Note that these values of A$_H$ are calculated 
from photometric observations, and not calculated from the A$_V$ values found 
as the best fit for spectral classification.

\item Compute the apparent bolometric magnitude $m_{bol}$ for each star and 
brown dwarf.  This can be done at any wavelength; here we use $H-$band.
The bolometric corrections are taken from a variety of sources in the 
literature for M dwarfs, corrected appropriately for filter transformations.  
For the M1-M4 and M6 spectral types, values from \citet[][see Table A5]{kh95} 
were used.  For M5, an average of values from \citet{kh95}, \citet{b91}, and 
\citet{l96} were used.  For M7, an interpolation of values given in 
\citet{kh95} and \citet{l96} was used.  And finally, for M8-M9, values from 
\citet{tmr93} were used.

\begin{equation}
m_{bol} = BC_H + H - A_H 
\end{equation}

\item Finally, $\log (L_{bol}/L_{\odot})$, the values that are plotted in Figure \ref{fig:hrdiagram}, can be computed:

\begin{equation}
\log \left(\frac{L_{bol}}{L_{\odot}}\right) = \frac{4.76 - m_{bol} + 5 \log d - 5}{2.5}
\end{equation}

\noindent
where $d =$ 450 pc.

\end{itemize}

\clearpage

\begin{deluxetable}{lcc}
\tablecaption{Summary of $Iz^{\prime}JHK$ Photometric Completeness}
\tabletypesize{\small}
\tablewidth{0pt}
\tablehead{
\colhead{Filter} & \colhead{Number of Sources \tablenotemark{a}} & 
\colhead{90\% Completeness Limits} \\
\colhead{\nodata} & \colhead{\nodata} & \colhead{[mag]}
}
\startdata
$I$ & 985 & 18.0\tablenotemark{b} \\
$z^{\prime}$ & 1092 & 18.1\tablenotemark{b} \\
$J$ & 1057 & 18.0 \\
$H$ & 1232 & 17.1 \\
$K$ & 1296 & 16.3
\enddata
\tablenotetext{a}{ Number of sources in the observed field with photometric 
uncertainties less than 0.15 magnitudes.}
\tablenotetext{b}{ The 90\% completeness limit quoted here is for the region 
north of a declination of $-5.15$; if we include the entire image in our 
analysis, the magnitude quoted is 80\% complete.}
\label{table:complete_summary}
\end{deluxetable}

\clearpage
\thispagestyle{empty}
\begin{deluxetable}{lcccccccllr}
\rotate
\tablecaption{Summary of Spectral Observations: Brown Dwarfs}
\tabletypesize{\scriptsize}
\tablewidth{0pt}
\tablehead{
\colhead{Name} & \colhead{R. A.\tablenotemark{a}} & 
\colhead{Dec.\tablenotemark{a}} & 
\colhead{I, ${\sigma}_I$\tablenotemark{b}}  & 
\colhead{z$^\prime$, ${\sigma}_{z^{\prime}}$\tablenotemark{b}} &
\colhead{J, ${\sigma}_J$\tablenotemark{c}} & 
\colhead{H, ${\sigma}_H$\tablenotemark{c}} & 
\colhead{K, ${\sigma}_K$\tablenotemark{c}} & 
\colhead{Sp. Type} &
\colhead{A$_H$\tablenotemark{i}} & \colhead{EW[H$\alpha$]}\\
\colhead{\nodata} & \colhead{(J2000)} & \colhead{(J2000)} & 
\colhead{(mag)} & \colhead{(mag)} & 
\colhead{(mag)} & \colhead{(mag)} & 
\colhead{(mag)} & \colhead{\nodata} &
\colhead{\nodata} & \colhead{(\AA)}
}
\startdata
1  & 05 35 27.4 & -05 09 04 & 17.882 $\pm$ 0.036 & 17.398 $\pm$ 0.042 & 14.852 $\pm$ 0.009 & 14.107 $\pm$ 0.008 & 13.594 $\pm$ 0.011 & M6.5\tablenotemark{d} & 0.2 & $-$19.2 \\
2  & 05 35 16.8 & -05 07 27 & 17.819 $\pm$ 0.018 & 16.903 $\pm$ 0.016 & 14.953 $\pm$ 0.009 & 14.413 $\pm$ 0.009 & 13.947 $\pm$ 0.011 & M6.5\tablenotemark{d} & 0.0 & $-$16.9 \\
3  & 05 35 13.0 & -05 02 09 & 17.393 $\pm$ 0.005 & 16.257 $\pm$ 0.008 & 14.465 $\pm$ 0.007 & 13.895 $\pm$ 0.007 & 13.477 $\pm$ 0.007 & M6.5\tablenotemark{d} & 0.0 & $-$25.8 \\
29 & 05 35 12.9 & -05 15 49 & 18.447 $\pm$ 0.051 & 17.441  $\pm$ 0.039 & 15.082 $\pm$ 0.015 & 14.309 $\pm$ 0.013 & 13.806 $\pm$ 0.015 & M8-M8.25\tablenotemark{e}, M8.5\tablenotemark{g} & 0.2 & neb\tablenotemark{j} \\
30 & 05 35 12.7 & -05 15 43 & 18.394 $\pm$ 0.069 & 17.378 $\pm$ 0.056 & 14.825 $\pm$ 0.010 & 13.951 $\pm$ 0.010 & 13.328 $\pm$ 0.008 & M7 \tablenotemark{e}, M7.25\tablenotemark{g} & 0.4 & neb\tablenotemark{j} \\
31 & 05 35 18.7 & -05 15 18 & 18.003 $\pm$ 0.062 & 16.369 $\pm$ 0.030 & 14.488 $\pm$ 0.009 & 13.658 $\pm$ 0.007 & 13.077 $\pm$ 0.009 & M7.75\tablenotemark{e}, M7.5\tablenotemark{h} & 0.3 & neb\tablenotemark{j} \\
21 & 05 35 38.2 & -05 03 34 & 18.561 $\pm$ 0.018 & 17.609 $\pm$ 0.017 & 16.112 $\pm$ 0.027 & 15.373 $\pm$ 0.031 & 14.840 $\pm$ 0.024 & M7\tablenotemark{f}, M7.25\tablenotemark{h} & 0.2 & $-$105.1 \\
34 & 05 35 13.6 & -05 14 22 & 18.445 $\pm$ 0.077 & 17.344 $\pm$ 0.051 & 15.793 $\pm$ 0.020 & 14.880 $\pm$ 0.017 & 14.335 $\pm$ 0.022 & M8\tablenotemark{f} & 0.4 & \nodata \\
77 & 05 35 40.3 & -05 12 32 & 19.290 $\pm$ 0.40? & 17.355 $\pm$ 0.098 & 15.462 $\pm$ 0.026 & 14.687 $\pm$ 0.018 & 14.155 $\pm$ 0.019 & M7\tablenotemark{f} & 0.2 & \nodata \\
923\_1621 & 05 35 15.1 & -04 58 09 & \nodata & \nodata & 17.867 $\pm$ 0.051 & 16.197 $\pm$ 0.028 & 15.200 $\pm$ 0.029 & M8\tablenotemark{g} & 1.5 & \nodata \\
1036\_108 & 05 35 10.1 & -05 14 56 & \nodata & \nodata & 16.551 $\pm$ 0.032 & 15.397 $\pm$ 0.022 & 14.594 $\pm$ 0.020 & M9\tablenotemark{g} & 0.8 & \nodata \\    
543\_681  & 05 35 32.0 & -05 08 35 & \nodata & \nodata & 16.848 $\pm$ 0.026 & 15.430 $\pm$ 0.017 & 14.485 $\pm$ 0.014 & M7.25\tablenotemark{g} & 1.2 & \nodata \\
766\_341  & 05 35 22.1 & -05 12 21 & \nodata & \nodata & 17.035 $\pm$ 0.081 & 15.908 $\pm$ 0.040 & 14.873 $\pm$ 0.027 & M7.75\tablenotemark{g} & 0.7 & \nodata \\
441\_326  & 05 35 36.6 & -05 12 31 & \nodata & \nodata & 16.547 $\pm$ 0.070 & 15.758 $\pm$ 0.058 & 14.906 $\pm$ 0.051 & M8.5\tablenotemark{g} & 0.3 & \nodata \\ 
988\_148  & 05 35 12.2 & -05 14 29 & \nodata & \nodata & 17.851 $\pm$ 0.069 & 16.503 $\pm$ 0.037 & 15.304 $\pm$ 0.030 & M8\tablenotemark{g} & 1.1 & \nodata \\
1004\_979 & 05 35 11.5 & -05 05 16 & \nodata & \nodata & 16.321 $\pm$ 0.019 & 14.986 $\pm$ 0.011 & 14.330 $\pm$ 0.011 & M7\tablenotemark{g} & 1.0 & \nodata\\
727\_1506 & 05 35 23.9 & -04 59 25 & \nodata & \nodata & 18.398 $\pm$ 0.070 & 16.800 $\pm$ 0.034 & 15.842 $\pm$ 0.031 & M7.75\tablenotemark{g} & 1.4 & \nodata\\
845\_236 & 05 35 23.3 & -05 13 48 & \nodata & \nodata & 16.714 $\pm$ 0.060 & 15.789 $\pm$ 0.037 & 14.607 $\pm$ 0.029 & M7\tablenotemark{g} & 0.5 & \nodata\\
735\_36 & 05 35 28.2 & -05 16 01 & \nodata & \nodata & 15.790 $\pm$ 0.049 & 14.673 $\pm$ 0.020 & 13.978 $\pm$ 0.021 & M8-M8.25\tablenotemark{g} & 0.7 & \nodata
\enddata
\tablenotetext{a}{ Coordinates from 2MASS final release.}
\tablenotetext{b}{ Magnitudes and relative uncertainties from 4-Shooter photometry.}
\tablenotetext{c}{ Magnitudes and relative uncertainties from SQIID photometry.}
\tablenotetext{d}{ Keck LRIS spectra.}
\tablenotetext{e}{ Blue Channel MMT spectra.}
\tablenotetext{f}{ Red Channel MMT spectra.}
\tablenotetext{g}{ SpeX IRTF spectra.}
\tablenotetext{h}{ CorMASS spectra.}
\tablenotetext{i}{ A$_H$ values calculated from $J-H$ colors (see Appendix \ref{sec:lbol_teff}) and not from the A$_V$ values found as the best fit from spectral classification.}
\tablenotetext{j}{ Too much emission from surrounding HII region to measure H$\alpha$.}
\label{table:browndwarfs}
\end{deluxetable}

\clearpage

\begin{deluxetable}{lcccccccllr}
\rotate
\tablecaption{Summary of Spectral Observations: Low Mass Members}
\tabletypesize{\scriptsize}
\tablewidth{0pt}
\tablehead{
\colhead{Name} & \colhead{R. A.\tablenotemark{a}} & 
\colhead{Dec.\tablenotemark{a}} & 
\colhead{I, ${\sigma}_I$\tablenotemark{b}} &
\colhead{z$^\prime$, ${\sigma}_{z^{\prime}}$\tablenotemark{b}} &
\colhead{J, ${\sigma}_J$\tablenotemark{c}} & 
\colhead{H, ${\sigma}_H$\tablenotemark{c}} & 
\colhead{K, ${\sigma}_K$\tablenotemark{c}} & 
\colhead{Sp. Type} & \colhead{A$_H$\tablenotemark{i}} & 
\colhead{EW[H$\alpha$]}\\
\colhead{\nodata} & \colhead{(J2000)} & \colhead{(J2000)} & 
\colhead{(mag)}& \colhead{(mag)}& 
\colhead{(mag)}& \colhead{(mag)} & 
\colhead{(mag)}& \colhead{\nodata}&
\colhead{\nodata}& \colhead{(\AA)}
}
\startdata
4  & 05 35 25.0 & -05 09 10 & 17.198 $\pm$ 0.039 & 16.374 $\pm$ 0.027 & 14.494 $\pm$ 0.007 & 13.904 $\pm$ 0.007 & 13.505 $\pm$ 0.014 & M6\tablenotemark{d} & 0.0 & $-$17.5 \\
5  & 05 35 24.1 & -05 09 07 & 16.444 $\pm$ 0.022 & 15.972 $\pm$ 0.025 & 14.374 $\pm$ 0.009 & 13.804 $\pm$ 0.007 & 13.385 $\pm$ 0.008 & M4.75\tablenotemark{d,g} & 0.0 & $-$32.6 \\
6.n & 05 35 16.2 & -05 09 19 & 14.853 $\pm$ 0.002 & 14.262 $\pm$ 0.003 & 12.916 $\pm$ 0.003 & 12.241 $\pm$ 0.003 & 11.996 $\pm$ 0.003 & M4.5-M4.75\tablenotemark{d} & 0.1 & $-$23.7 \\
13 & 05 35 12.7 & -05 12 01 & 16.471 $\pm$ 0.008 & 15.655 $\pm$ 0.005 & 14.164 $\pm$ 0.006 & 13.535 $\pm$ 0.006 & 13.243 $\pm$ 0.007 & M5.5\tablenotemark{e} & 0.0 & \nodata \\
15 & 05 35 35.7 & -05 10 51 & 16.360 $\pm$ 0.004 & 15.633 $\pm$ 0.005 & 14.380 $\pm$ 0.007 & 13.768 $\pm$ 0.006 & 13.381 $\pm$ 0.008 & M5.5\tablenotemark{e} & 0.0 & \nodata \\
17 & 05 35 42.7 & -05 10 16 & 17.234 $\pm$ 0.030 & 16.393 $\pm$ 0.030 & 14.833 $\pm$ 0.010 & 14.233 $\pm$ 0.010 & 13.784 $\pm$ 0.011 & M5.5\tablenotemark{f} & 0.0 & $-$36.4 \\
18 & 05 35 15.6 & -05 09 32 & 16.114 $\pm$ 0.004 & 15.411 $\pm$ 0.005 & 13.943 $\pm$ 0.006 & 13.340 $\pm$ 0.005 & 13.023 $\pm$ 0.006 & M5.75\tablenotemark{e} & 0.0 & \nodata \\
20 & 05 35 37.0 & -05 05 26 & 17.164 $\pm$ 0.024 & 16.247 $\pm$ 0.015 & 14.702 $\pm$ 0.008 & 14.128 $\pm$ 0.007 & 13.701 $\pm$ 0.008 & M6-M6.25\tablenotemark{e} & 0.0 & \nodata \\
23 & 05 35 40.0 & -05 02 37 & 16.239 $\pm$ 0.002 & 15.482 $\pm$ 0.003 & 14.111 $\pm$ 0.006 & 13.489 $\pm$ 0.006 & 13.159 $\pm$ 0.006 & M5.5\tablenotemark{e} & 0.0 & $-$16.4 \\
24 & 05 35 19.7 & -05 02 29 & 16.927 $\pm$ 0.007 & 16.088 $\pm$ 0.011 & 14.502 $\pm$ 0.007 & 13.882 $\pm$ 0.007 & 13.524 $\pm$ 0.007 & M5.75\tablenotemark{e} & 0.0 & \nodata \\
27 & 05 34 43.5 & -05 14 43 & 16.889 $\pm$ 0.015 & 16.189 $\pm$ 0.011 & 14.894 $\pm$ 0.009 & 14.115 $\pm$ 0.008 & 13.697 $\pm$ 0.008 & M5\tablenotemark{e} & 0.3 & \nodata \\
33 & 05 35 16.8 & -05 14 47 & 18.041 $\pm$ 0.079 & 16.855 $\pm$ 0.040 & 15.257 $\pm$ 0.013 & 14.255 $\pm$ 0.010 & 13.636 $\pm$ 0.012 & M5.5\tablenotemark{f} & 0.6 & \nodata \\
70 & 05 35 16.3 & -05 04 36 & 16.805 $\pm$ 0.005 & 16.061 $\pm$ 0.012 & 14.621 $\pm$ 0.008 & 13.974 $\pm$ 0.007 & 13.729 $\pm$ 0.008 & M5.5\tablenotemark{e} & 0.1 & \nodata \\
75 & 05 34 44.9 & -05 12 32 & 18.732 $\pm$ 0.039 & 17.419 $\pm$ 0.033 & 15.324 $\pm$ 0.014 & 14.327 $\pm$ 0.012 & 13.752 $\pm$ 0.014 & M6\tablenotemark{f}, M5.75\tablenotemark{g} & 0.6 & \nodata \\
76 & 05 35 20.2 & -05 12 11 & 16.623 $\pm$ 0.073 & 15.660 $\pm$ 0.036 & 14.978 $\pm$ 0.010 & 14.102 $\pm$ 0.009 & 13.630 $\pm$ 0.009 & M5.5\tablenotemark{f} & 0.4 & \nodata \\
78 & 05 34 56.7 & -05 13 57 & 19.444 $\pm$ 0.209 & 18.170 $\pm$ 0.112 & 15.901 $\pm$ 0.015 & 14.805 $\pm$ 0.011 & 14.297 $\pm$ 0.011 & M4.5\tablenotemark{f} & 0.7 & \nodata \\
636\_972 & 05 35 32.6 & -05 05 38 & 18.011 $\pm$ 0.032 & 17.174 $\pm$ 0.034 & 15.805 $\pm$ 0.014 & 14.791 $\pm$ 0.010 & 14.164 $\pm$ 0.011 & M4.25\tablenotemark{g} & 0.6 & \nodata \\
369\_674 & 05 35 44.5 & -05 08 56 & \nodata  & 19.125 $\pm$ 0.251 & 16.149 $\pm$  0.021 & 14.831 $\pm$   0.012 &  14.073 $\pm$   0.011 & M4.25\tablenotemark{g} & 1.0 & \nodata \\
774\_120 & 05 35 26.5 & -05 15 05 & \nodata & \nodata & \nodata & 14.909 $\pm$ 0.019 & 14.174 $\pm$ 0.018 & M5.5\tablenotemark{g} & \nodata & \nodata \\
1213\_413 & 05 35 06.9 & -05 11 50 & \nodata & \nodata & 16.702 $\pm$ 0.043 & 15.601 $\pm$ 0.030 & 14.928 $\pm$ 0.024 & M4-M4.5\tablenotemark{g} & 0.7 & \nodata \\
CHS 8782 & 05 35 16.2 & -05 15 48 & 17.157 $\pm$ 0.013 & 16.448 $\pm$ 0.023 &  14.935 $\pm$ 0.011 & 14.262 $\pm$ 0.011 & 13.869 $\pm$ 0.014 & M5.25\tablenotemark{h} & 0.1 & \nodata \\
\enddata
\tablenotetext{a}{ Coordinates from 2MASS final release.}
\tablenotetext{b}{ Magnitudes and relative uncertainties from 4-Shooter photometry.}
\tablenotetext{c}{ Magnitudes and relative uncertainties from SQIID photometry.}
\tablenotetext{d}{ Keck LRIS spectra.}
\tablenotetext{e}{ Blue Channel MMT spectra.}
\tablenotetext{f}{ Red Channel MMT spectra.}
\tablenotetext{g}{ SpeX IRTF spectra.}
\tablenotetext{h}{ CorMASS spectra.}
\tablenotetext{i}{ A$_H$ values calculated from $J-H$ colors (see Appendix \ref{sec:lbol_teff}) and not from the A$_V$ values found as the best fit from spectral classification.}
\label{table:lowmassmemberspectra}
\end{deluxetable}

\clearpage

\begin{deluxetable}{lcccccccll}
\rotate
\tablecaption{Summary of Spectral Observations: Field Stars and Objects of Indeterminate Type}
\tabletypesize{\scriptsize}
\tablewidth{0pt}
\tablehead{
\colhead{Name} & \colhead{R. A.\tablenotemark{a}} & 
\colhead{Dec.\tablenotemark{a}} & 
\colhead{I, ${\sigma}_I$\tablenotemark{b}} & 
\colhead{z$^\prime$, ${\sigma}_{z^{\prime}}$\tablenotemark{b}} &
\colhead{J, ${\sigma}_J$\tablenotemark{c}} & 
\colhead{H, ${\sigma}_H$\tablenotemark{c}} & 
\colhead{K, ${\sigma}_K$\tablenotemark{c}} & 
\colhead{Sp. Type} & \colhead{A$_H$\tablenotemark{i}}\\
\colhead{\nodata} & \colhead{(J2000)} & \colhead{(J2000)} & 
\colhead{(mag)} & \colhead{(mag)} & 
\colhead{(mag)} & \colhead{(mag)} & 
\colhead{(mag)} & \colhead{\nodata} & \colhead{\nodata}
}
\startdata
6  & 05 35 15.5 & -05 08 59 & 17.332 $\pm$ 0.010 & 16.722 $\pm$ 0.022 & 15.716 $\pm$ 0.013 & 15.078 $\pm$ 0.013 & 14.841 $\pm$ 0.015 & M3.25V\tablenotemark{d} & 0.1 \\
12 & 05 34 54.9 & -05 02 43 & 19.555 $\pm$ 0.042 & 18.293 $\pm$ 0.052 & 17.084 $\pm$ 0.029 & 16.052 $\pm$ 0.019 & 15.631 $\pm$ 0.021 & \nodata\tablenotemark{f} & 0.6 \\
16 & 05 35 09.0 & -05 10 26 & 15.938 $\pm$ 0.003 & 15.579 $\pm$ 0.005 & 14.602 $\pm$ 0.008 & 13.926 $\pm$ 0.008 & 13.711 $\pm$ 0.010 & M2.5V\tablenotemark{e} & 0.1 \\
25 & 05 35 07.8 & -05 01 19 & 19.655 $\pm$ 0.070 & 18.316 $\pm$ 0.055 & 16.254 $\pm$ 0.018 & 15.463 $\pm$ 0.016 & 15.003 $\pm$ 0.016 & M8IV\tablenotemark{f,k} & 0.3\\
01 & 05 34 46.8 & -05 04 27 & 19.837 $\pm$ 0.062 & 18.247 $\pm$ 0.046 & 16.910 $\pm$ 0.027 & 15.849 $\pm$ 0.018 & 15.335 $\pm$ 0.019 & M4V\tablenotemark{f} & 0.7 \\
74 & 05 34 58.2 & -05 07 13 & 20.022 $\pm$ 0.111 & 18.542 $\pm$ 0.072 & 16.868 $\pm$ 0.023 & 15.729 $\pm$ 0.017 & 15.213 $\pm$ 0.017 & M3V\tablenotemark{f} & 0.8 \\
79 & 05 34 43.5 & -05 14 21 & 17.544 $\pm$ 0.024 & 17.035 $\pm$ 0.023 & 15.300 $\pm$ 0.014 & 14.287 $\pm$ 0.009 & 13.976 $\pm$ 0.010 &K?\tablenotemark{f} & 0.6 \\
558\_371  & 05 35 31.4 & -05 12 00 & \nodata & \nodata & 14.558 $\pm$ 0.010 & 13.962 $\pm$ 0.011 & 13.656 $\pm$ 0.013 & M5V\tablenotemark{g} & 0.0 \\
650\_963  & 05 35 27.3 & -05 05 27 & \nodata & \nodata & 17.220 $\pm$ 0.034 & 16.340 $\pm$ 0.023 & 15.601 $\pm$ 0.028 & \nodata\tablenotemark{g} & 0.4 \\
820\_1131 & 05 35 19.7 & -05 03 35 & \nodata & \nodata & \nodata &  16.705 $\pm$ 0.034 & 15.270 $\pm$ 0.017 & \nodata\tablenotemark{g} & \nodata \\
513\_395  & 05 35 33.4 & -05 11 45 & \nodata & \nodata & 16.899 $\pm$ 0.070 & 16.012 $\pm$ 0.073 & 14.909 $\pm$ 0.053 & \nodata\tablenotemark{g} & 0.4 \\
1107\_414 & 05 35 06.9 & -05 11 33 & \nodata & \nodata & 17.877 $\pm$ 0.050 & 16.136 $\pm$ 0.028 & 15.145 $\pm$ 0.021 & \nodata\tablenotemark{g} & 1.6 \\
652\_1582 & 05 35 27.2 & -04 58 35 & \nodata & \nodata & \nodata & 17.215 $\pm$ 0.057 & 15.571 $\pm$ 0.027 & \nodata\tablenotemark{g} & \nodata \\
1164\_263 & 05 35 04.3 & -05 13 13 & \nodata & \nodata & 17.380 $\pm$ 0.047 & 16.251 $\pm$ 0.031 & 15.317 $\pm$ 0.027 & \nodata\tablenotemark{g} & 0.7 \\
455\_461  & 05 35 36.0 & -05 11 01 & \nodata & \nodata & \nodata & 16.854 $\pm$ 0.063 & 15.692 $\pm$ 0.048 & \nodata\tablenotemark{g} & \nodata \\
995\_1280 & 05 35 11.9 & -05 01 56 & \nodata & \nodata & 17.865 $\pm$ 0.042 & 16.292 $\pm$ 0.024 & 15.263 $\pm$ 0.017 & \nodata\tablenotemark{g} & 1.4 \\
1095\_766 & 05 35 07.4 & -05 07 38 & \nodata & \nodata & 17.661 $\pm$ 0.039 & 15.730 $\pm$ 0.017 & 14.665 $\pm$ 0.013 & \nodata\tablenotemark{g} & 1.9 \\
1068\_598 & 05 35 08.6 & -05 09 30 & \nodata & \nodata & 17.117 $\pm$ 0.034 & 15.550 $\pm$ 0.016 & 14.728 $\pm$ 0.013 & \nodata\tablenotemark{g} & 1.4 \\
925\_1442 & 05 35 15.0 & -05 00 08 & \nodata & \nodata & \nodata & 16.350 $\pm$ 0.052 & 14.827 $\pm$ 0.023 & \nodata\tablenotemark{g} & \nodata \\
657\_1084 & 05 35 27.0 & -05 04 07 & \nodata & \nodata & \nodata & \nodata & 14.077 $\pm$ 0.033 & \nodata\tablenotemark{g} & \nodata \\
814\_1360 & 05 35 20.0 & -05 01 03 & \nodata & \nodata & \nodata & \nodata & 14.325 $\pm$ 0.019 & \nodata\tablenotemark{g} & \nodata \\
643\_957 & 05 35 27.6 & -05 05 31 & \nodata & \nodata & \nodata & \nodata & 14.870 $\pm$ 0.024 & \nodata\tablenotemark{g} & \nodata \\
706\_829 & 05 35 24.8 & -05 06 56 & \nodata & \nodata & \nodata & \nodata & 13.897 $\pm$ 0.009 & \nodata\tablenotemark{g} & \nodata \\
356\_225 & 05 35 45.1 & -05 13 56 & 18.086 $\pm$ 0.066 & 17.684 $\pm$ 0.071 & 16.479 $\pm$ 0.030 & 15.397 $\pm$ 0.026 & 14.776 $\pm$ 0.029 & M4V\tablenotemark{g} & 0.7 \\
1191\_569 & 05 35 07.9 & -05 10 06 & \nodata & \nodata & 16.922 $\pm$ 0.026 & 15.534 $\pm$ 0.016 & 14.893 $\pm$ 0.014 & \nodata\tablenotemark{g} & 1.1 \\
1163\_564 & 05 35 09.1 & -05 10 09 & \nodata & \nodata & 16.993 $\pm$ 0.040 & 15.699 $\pm$ 0.019 & 14.995 $\pm$ 0.019 & \nodata\tablenotemark{g} & 1.0 \\
469\_431 & 05 35 40.0 & -05 11 38 & 18.377 $\pm$ 0.071 & 17.913 $\pm$ 0.088 & 16.568 $\pm$ 0.029 & 15.473 $\pm$ 0.023 & 14.830 $\pm$ 0.019 & \nodata\tablenotemark{g} & 0.7 \\
1448\_85 & 05 34 56.4 & -05 15 28 & \nodata & \nodata & 17.357 $\pm$ 0.055 & 15.817 $\pm$ 0.027 & 14.915 $\pm$ 0.029 & \nodata\tablenotemark{g} & 1.3 \\
391\_684 & 05 35 43.5 & -05 08 50 & 18.052 $\pm$ 0.056 & 17.801 $\pm$ 0.036 & 15.790 $\pm$ 0.016 & 14.777 $\pm$ 0.011 & 14.091 $\pm$ 0.010 & \nodata\tablenotemark{g} & 0.6 \\
708\_1277 & 05 35 29.4 & -05 02 15 & \nodata & \nodata & 17.118 $\pm$ 0.025 & 15.490 $\pm$ 0.014 & 14.710 $\pm$ 0.014 & \nodata\tablenotemark{g} & 1.4 \\
1821\_1428 & 05 34 39.8 & -05 00 34 & 18.822 $\pm$ 0.027 & 18.317 $\pm$ 0.043 & 16.723 $\pm$ 0.027 & 15.508 $\pm$ 0.019 & 14.813 $\pm$ 0.018 & \nodata\tablenotemark{g} & 0.9 \\
1350\_447 & 05 35 00.8 & -05 11 27 & \nodata & \nodata & 16.772 $\pm$ 0.026 & 15.193 $\pm$ 0.014 & 14.440 $\pm$ 0.012 & \nodata\tablenotemark{g} & 1.4 \\
1137\_1202 & 05 35 10.3 & -05 03 05 & \nodata & \nodata & 17.001 $\pm$ 0.032 & 15.128 $\pm$ 0.013 & 14.183 $\pm$ 0.010 & \nodata\tablenotemark{g} & 1.8 \\
558\_282 & 05 35 36.1 & -05 13 17 & \nodata & \nodata & 16.714 $\pm$ 0.060 & 15.027 $\pm$ 0.029 & 14.217 $\pm$ 0.031 & \nodata\tablenotemark{g} & 1.5 \\
554\_173 & 05 35 36.3 & -05 14 30 & \nodata & \nodata & \nodata & 14.455 $\pm$ 0.014 & 13.818 $\pm$ 0.013 & \nodata\tablenotemark{g} & \nodata \\
1442\_484 & 05 34 56.7 & -05 11 03 & \nodata & \nodata & 17.141 $\pm$ 0.032 & 15.657 $\pm$ 0.021 & 14.941 $\pm$ 0.017 & \nodata\tablenotemark{g} & 1.2 \\
1526\_99 & 05 34 53.0 & -05 15 19 & \nodata & \nodata & 16.288 $\pm$ 0.022 & 15.000 $\pm$ 0.014 & 14.404 $\pm$ 0.014 & \nodata\tablenotemark{g} & 1.0 \\
164\_563(a,b)\tablenotemark{j} & 05 35 53.6 & -05 10 10 & 18.081 $\pm$ 0.028 & 17.667 $\pm$ 0.040 & 16.185 $\pm$ 0.018 & 15.237 $\pm$ 0.015 & 14.681 $\pm$ 0.015 & \nodata\tablenotemark{g} & 0.5 \\
458\_1202b & 05 35 40.5 & -05 03 05 & 19.173 $\pm$ 0.044 & 18.615 $\pm$ 0.043 & 16.610 $\pm$ 0.038 & 15.278 $\pm$ 0.016 & 14.706 $\pm$ 0.016 & \nodata\tablenotemark{g} & 1.0 \\
458\_1202a & 05 35 40.3 & -05 03 06 & 19.308 $\pm$ 0.031 & 18.721 $\pm$ 0.044  & 17.194 $\pm$ 0.054 & 15.783 $\pm$ 0.028 & 15.290 $\pm$ 0.030 & \nodata\tablenotemark{g} & 1.1 \\
269\_588 & 05 35 48.9 & -05 09 53 & 19.078 $\pm$ 0.032 & 17.509 $\pm$ 0.075 & 16.243 $\pm$ 0.018 & 15.069 $\pm$ 0.013 & 14.513 $\pm$ 0.016 & \nodata\tablenotemark{g} & 0.8 \\
781\_708 & 05 35 26.2 & -05 08 34 & \nodata & \nodata & 15.490 $\pm$ 0.026 & 15.200 $\pm$ 0.027 & 14.607 $\pm$ 0.023 & \nodata\tablenotemark{g} & 0.0 \\
301\_833 & 05 35 47.5 & -05 07 10 & 18.363 $\pm$ 0.067 & \nodata & 16.308 $\pm$ 0.020 & 15.212 $\pm$ 0.013 & 14.672 $\pm$ 0.013 & \nodata\tablenotemark{g} & 0.7 \\
1224\_1205 & 05 35 06.4 & -05 03 03 & 19.527 $\pm$ 0.033 & 18.321 $\pm$ 0.024 & 16.191 $\pm$ 0.016 & 14.859 $\pm$ 0.011 & 14.289 $\pm$ 0.011 & \nodata\tablenotemark{g} & 1.0 \\
1512\_410 & 05 34 53.6 & -05 11 52 & 19.043 $\pm$ 0.040 & 17.962 $\pm$ 0.040 & 15.698 $\pm$ 0.013 & 14.480 $\pm$ 0.010 & 13.897 $\pm$ 0.009 & \nodata\tablenotemark{g} & 0.9 \\
583\_352 & 05 35 35.0 & -05 12 31 & \nodata & \nodata & 16.405 $\pm$ 0.062 & 15.473 $\pm$ 0.057 & 14.568 $\pm$ 0.095 & \nodata\tablenotemark{h} & 0.5 \\
d216-0939 & 05 35 21.6 & -05 09 39 & \nodata & \nodata & 13.879 $\pm$ 0.006 & 12.918 $\pm$ 0.005 & 12.222 $\pm$ 0.004 & $<$ M3\tablenotemark{h} & 0.5 \\
CHS 9695 & 05 35 22.4 & -05 07 39 & \nodata & \nodata & 14.741 $\pm$ 0.014 & 12.087 $\pm$ 0.004 & 10.268 $\pm$ 0.002 & $\lae$ M3\tablenotemark{h} & 2.9 \\
\enddata
\tablenotetext{a}{ Coordinates from 2MASS final release.}
\tablenotetext{b}{ Magnitudes and relative uncertainties from 4-Shooter photometry.}
\tablenotetext{c}{ Magnitudes and relative uncertainties from SQIID photometry.}
\tablenotetext{d}{ Keck LRIS spectra.}
\tablenotetext{e}{ Blue Channel MMT spectra.}
\tablenotetext{f}{ Red Channel MMT spectra.}
\tablenotetext{g}{ SpeX IRTF spectra.}
\tablenotetext{h}{ CorMASS spectra.}
\tablenotetext{i}{ A$_H$ values calculated from $J-H$ colors (see Appendix \ref{sec:lbol_teff}) and not from the A$_V$ values found as the best fit from spectral classification.}
\tablenotetext{j}{ Source is unresolved in imaging but was resolved on the slit.}
\tablenotetext{k}{ This source has a spectral type later than M6.5 and is a young brown dwarf in OMC 2/3.  However, it serendipitously fell into the slit during the observation and although we have reconstructed which source it is, and believe it is the one whose coordinates and magnitudes are listed here, until we are confident this is the correct source, we have chosen to put it in this table.  It is also an accreting brown dwarf; it has an EW[H$\alpha$] $<$ $-$40.}
\label{table:fieldspectra}
\end{deluxetable}

\clearpage

\clearpage

\clearpage

\clearpage

\clearpage

\clearpage

\clearpage

\clearpage

\clearpage

\clearpage

\clearpage

\clearpage

\clearpage

\clearpage

\clearpage

\clearpage

\clearpage

\end{document}